\documentclass[%
 pra,reprint,
 amsmath,amssymb,
 aps,
superscriptaddress
]{revtex4-2}

\usepackage{graphicx}
\usepackage{dcolumn}
\usepackage{bm}
\usepackage{hyperref}
\usepackage[mathlines]{lineno}
\usepackage{physics}
\usepackage{lipsum} 
\usepackage[caption=false]{subfig}
\usepackage{float}

\newcommand{\eref}[1]{(\ref{#1})}

\usepackage{tikz-cd}
\usetikzlibrary{decorations.pathmorphing}
\usetikzlibrary{arrows,
    chains,
    decorations.markings,
    shadows, shapes.arrows,shapes, fit}
    
\tikzstyle{tensor}=[rectangle,draw=blue!50,fill=blue!20,thick]
\tikzset{%
dotted_block/.style={draw=black!20!white, line width=1pt, dash pattern=on 3pt off 1pt,
            inner ysep=3mm,inner xsep=2mm, rectangle, rounded corners}
}

\begin{document}


\title{Thermal cycle and polaron formation in structured bosonic environments}

\author{Angela Riva}
\affiliation{LPENS, Département de physique, Ecole normale supérieure, Centre Automatique et Systèmes (CAS), MINES ParisTech, Université PSL, Sorbonne Université, CNRS, Inria, 75005 Paris, France}
\affiliation{Sorbonne Universit\'{e}, CNRS, Institut des NanoSciences de Paris, 4 place Jussieu, 75005 Paris, France}
\affiliation{Dipartimento di Fisica ``Aldo Pontremoli'', Universit{\`a} degli Studi di Milano, Via Celoria 16, 20133 Milano-Italy}
\author{Dario Tamascelli}
\affiliation{Dipartimento di Fisica ``Aldo Pontremoli'', Universit{\`a} degli Studi di Milano, Via Celoria 16, 20133 Milano-Italy}
\affiliation{Institut f\"ur Theoretische Physik und IQST, Albert-Einstein-Allee 11, Universit\"at Ulm, D-89081 Ulm, Germany}
\author{Angus J.~Dunnett}
\affiliation{Sorbonne Universit\'{e}, CNRS, Institut des NanoSciences de Paris, 4 place Jussieu, 75005 Paris, France}
\author{Alex W.~Chin}
\affiliation{Sorbonne Universit\'{e}, CNRS, Institut des NanoSciences de Paris, 4 place Jussieu, 75005 Paris, France}



\begin{abstract}
Chain-mapping techniques combined with the time-dependent density matrix renormalization group are powerful tools for simulating the dynamics of open quantum systems interacting with structured bosonic environments. Most interestingly, they leave the degrees of freedom of the environment open to inspection. In this work, we fully exploit the access to environmental observables to illustrate how the evolution of the open quantum system can be related to the detailed evolution of the environment it interacts with. In particular, we give a precise description of the fundamental physics that enables the finite temperature chain-mapping formalism to express dynamical equilibrium states. Furthermore, we analyze a two-level system strongly interacting with a super-Ohmic environment, where we discover a change in the spin-boson ground state that can be traced to the formation of polaronic states.
\end{abstract}

\maketitle

\section{\label{sec:intro}Introduction}

\begin{figure*}
\includegraphics[width=0.8\textwidth]{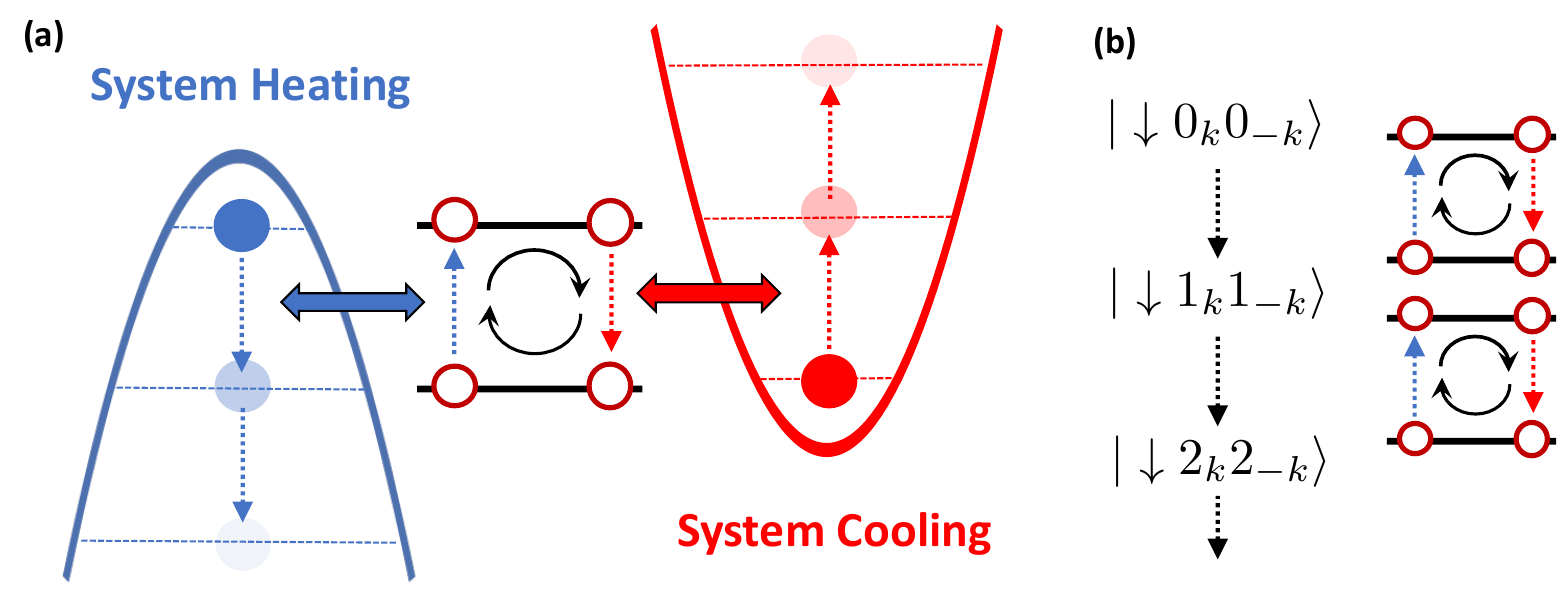}
\caption{\label{cycle} The heating cycle. (a) In the extended environment, excitation of the two-level system is accompanied by creation of an excitation in the resonant negative frequency mode (system heating). Once excited, the system can decay back to the ground state via creation of an excitation in the positive frequncy mode (the direction of the cycle shown is arbitrary). The ratio of the uphill and downhill transition rates obeys detailed balance and thus maintains the correct (mean) thermal populations whilst allowing for thermal fluctuations. Due to the unbounded Fock spaces of the bosonic modes, this process can continue indefinitely, leading to an unbounded growth of environmental excitations.  (b) As excitation in the environment cannot be destroyed, one pair of extra excitations must be created per cycle, leading to environmental wave functions developing pair-like quantum correlations among the positive and negative resonant modes.}
\end{figure*}

The theory of open quantum systems (OQS) provides the fundamental description of how irreversible and noisy processes emerge when a quantum system interacts with the continua of excitations constituting its surrounding `environment' \cite{breuer_petruccione,rivas2012open,Leggett}. As these processes include the ubiquitous - and essentially unavoidable -  real-world phenomena of energy relaxation, dephasing and decoherence, the physics of OQS are implicated in a vast range of quantum phenomena across physics, chemistry and biology, and are especially important for future technologies aiming to exploit quantum effects for computing, communication and energy applications \cite{rotter2015review,koch2022quantum,auffeves2022quantum,farina2019charger,wertnik2018optimizing}.  

In recent years, particular interest has arisen in the physics of OQS in the presence of strong system-environment interactions and persistent (non-Markovian) system-environment memory effects, both of which are typical of functional molecular and biological nanomaterials \cite{alvertis2020impact,dunnett2021influence}. Under these conditions, the dissipative dynamics can no longer be described by simple master equations, as the system and its environments both evolve, and mutually influence each other's evolution, on timescales relevant to the process under study. The correct description must therefore retain real-time details of the state of the system, its many body environment \emph{and} the correlations that arise between them, which presents a daunting computational challenge.

Tensor network methods \cite{Schollwoeck_2011,Orus_2014}, and in particular efficient time evolution algorithms for Matrix Product States (MPS) \cite{vidal03,vidal04,daley04,zwolak04,haegeman_unifying,Lubich_Oseledets_Vandereycken_2015}, have recently opened up the possibility of doing precisely this, allowing for a microscopic characterization of non-Markovian reduced state dynamics by representing and propagating the full wave function of the system and its surrounding environment with numerical exactitude. The power and versatility of this approach, in the form of the original Time Evolving Density operator with Orthogonal Polynomials Algorithm (TEDOPA), have been demonstrated through multiple applications in photonic, molecular and fundamental models of decoherence \cite{prior2010efficient,oviedo2016phase,del2018tensor}; but, as a pure wave function method, the majority of these studies were confined to zero-temperature environments. Unfortunately, extending this formulism to mixed environment states at finite temperatures would, \textit{a priori}, require thermal averaging over many expensive simulation runs, and would quickly become intractable, even at fairly low temperatures, due to the rapid proliferation of environmental state configurations.        

However, the recent development of the Thermalized-TEDOPA (T-TEDOPA) technique appears to have effectively and elegantly resolved this issue, enabling us to recover finite temperature results from the evolution of a single pure initial state \cite{Tamascelli_ttedopa, TAMA2020}. As will be described in more detail below, the essential idea is to introduce new environmental modes with negative frequencies such that, from the point of view of the reduced system dynamics, the time evolution starting from the vacuum state of such an extended environment is equivalent to the one that would be obtained by starting from a thermal state in the original environment. A rigorous theorem due to Tamascelli {\it et al.} guarantees that such an equivalent system can always be found, and also how it is to be constructed \cite{Tamascelli_ttedopa}.  As pure states admit a straightforward MPS representation that can be efficiently time-evolved, T-TEDOPA seemingly presents remarkably advantageous computational performance compared to sampling and/or costly evolution of density matrix operators, and has been shown to correctly capture finite temperature physics in a range of physical settings that permit comparisons with real experimental data \cite{dunnett2021influence,Dunnett_Chin_2021_evolving}.

In this paper we comprehensively explore the fundamental system-environment physics that enables T-TEDOPA to efficiently `mimic' finite-temperature effects, and critically asses the true numerical advantage that can be found by the combining T-TEDOPA with recent MPS propagation techniques. Indeed, it is generally observed - though not fully explained - that T-TEDOPA simulations require more numerical resources, i.e. higher bond and local Hilbert space dimensions, as well as chain lengths \cite{Dunnett_Chin_2021_vibr,Dunnett_Chin_2021_evolving}.  This work is motivated by a possible underlying reason for this: the recent observation in Ref. \cite{Dunnett_Chin_2021} of an unbounded growth in the total number of excitations in the extended environment for all finite temperatures. This continuous growth of excitations in the initially empty environment was shown to occur even after the system observables had completely relaxed to their (thermal) steady state values, raising concerns about the numerical efficiency of T-TEDOPA for long-time simulations.   

\begin{figure*}
\includegraphics[width=0.8\textwidth]{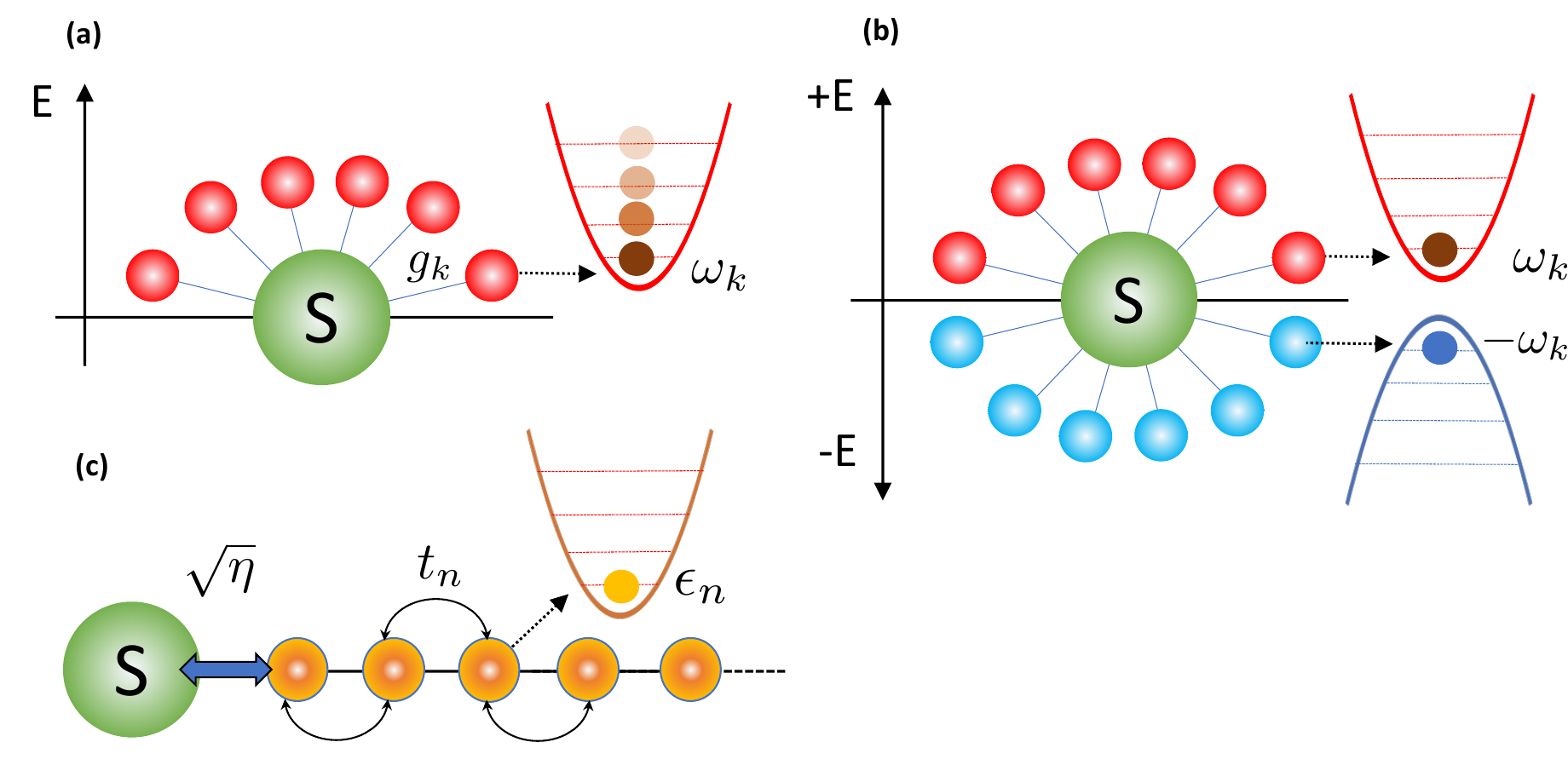}
\caption{\label{environ} Three equivalent representations of a system coupled to an environment that are compared in this work. (a) a standard, physical representation of a system coupled to a continuum of oscillators, each of which is in a thermal Gibbs state at temperature $\beta^{-1}$. (b) The  extended environment reproduces the effect of the environment of (a) by introducing a second environment containing modes with negative frequencies and a new, temperature depedent spectral density $J(\omega,\beta)$ describing the couplings of positive and frequency modes to the system. Crucially, the initial state of this extended environment that reproduces the physics of (a) is the vacuum.  (c) The T-TEDOPA nearest-neighbour chain representation is obtained from (b) via a unitary transform based on the $J(\omega,\beta)$. This $1$D form of the Hamiltonian can be efficiently simulated with MPS methods and provides complementary insight into the environment dynamics we present for (a) and (b).    }
\end{figure*}

A tentative explanation, referred to as the `heating cycle' instability, was proposed in Ref. \cite{Dunnett_Chin_2021}, and is sketched in Fig. \ref{cycle}. This posits that bath dynamics must continue, even after system relaxation, in order to capture the thermal fluctuations characteristic of the thermal state. For a simple two-level system interacting with a bath in the extended representation, energetically uphill (downhill) transitions are associated with the \emph{creation} of resonant negative(positive)-energy bath excitations. The fact that \emph{both} system excitation and de-excitation can be considered as emission into the positive/negative frequency environments, means that both processes can occur spontaneously, as a result of bosonic quantum fluctuations. This could thus create a continuous creation of excitations peaked around the positive and negative resonant frequencies of the extended environment. Unfortunately, the study in Ref. \cite{Dunnett_Chin_2021} only considered Ohmic environments, and the expected resonant peaks in the extended environment populations could not be resolved due to the additional presence of thermal pure dephasing processes which cause a large and broad growth of excitations around zero frequency that masks all other features.            

A first objective of this work is to complete the description of how dynamical equilibrium states are expressed in the T-TEDOPA formalism, by obtaining conclusive evidence for the onset of the thermal cycle instability. Here we confirm the hypothesis of Ref. \cite{Dunnett_Chin_2021} in two cases: (1) super-Ohmic environments where low frequency noise is suppressed and (2) - to our initial surprise - sub-Ohmic environments that have strongly \emph{enhanced} low-frequency couplings. Exploiting the access to full many body information in our T-TEDOPA/MPS approach, we explore how this apparent instability manifests in the chain representation used for simulations ( Fig. \ref{environ}), explicitly connecting the unbounded growth of excitations in the extended environment to the greater numerical resources needed for finite-temperature simulations. Crucially, this analysis shows that there are no pathologies in the chain representation of the dynamics:  excitation populations on the \emph{local} chain oscillators are always bounded, so T-TEDOPA simulations can always give controllably accurate results for arbitrarily long times and temperatures. 

A further implication of the heating cycle is that the creation of positive and negative frequency excitations occurs in pairs, so that quantum pair correlations should spontaneously develop during the dynamics (Fig. \ref{cycle}). Here, we explicitly reveal the existence of such correlations in the extended environments, and relate these back to the thermofield approach of De Vega and Bañuls which is based on the use of two-mode squeezed vacuum states to mimic the effect of a mixed thermal environment \cite{devegaban}. Indeed, by inverting the thermofield transformation, we are finally able to ``close'' the triangle of environment representations shown in Fig. \ref{environ}, allowing the numerical results in the T-TEDOPA chain and the underlying extended environment to be expressed in the \emph{original} (physical) thermal environment. Analysis in this representation at high and low temperatures provides insights into fundamental questions such as when an environment can be considered as a Markovian heat bath, which we shall briefly discuss.         

Finally, while most of our numerical experiments are conducted under conditions where a two-level system interacts with the thermal environment through simple and incoherent energy exchange (uphill and downhill transitions), the non-perturbative nature of T-TEDOPA also permits exploration of situations when this picture breaks down. In this work, we discover and analyze such a case, where a non-perturbative coupling of a TLS to high frequency super-Ohmic environments causes a change in the global ground state of $S+E$, leading to the spontanteous emergence of coherence and weakly damped oscillatory motion. The dynamics in this regime exhibit non-Markovian features that cannot be captured without treating at least some of the environmental degrees of freedom on the same footing of the system \cite{tamaSmirne22,closure}, and the application of our analysis framework - primarily our measure of quantum correlations in the extended environment - allows us to trace the observed behavior to the rapid formation of polaronic states, followed by slow, weakly damped tunneling of the polarons between degenerate configurations of the environment in the new ground state.    

The rest of the article is organized as follows: In Sec. \ref{sec:methods} we briefly describe the usage of the T-TEDOPA chain mapping. In Sec.~\ref{sec:results} we provide details on the setting used in our investigation. Section~\ref{sec:pure_dephasing} is devoted the discussion of how low frequency noise due to pure dephasing dominates in the sub-Ohmic environment.  We then examine the environmental correlations as signature of the onset of a thermal cycle in the presence of super-Ohmic environments in Sec. \ref{sec:cycle}. Finally, in Sec.~\ref{sec:s=3} we investigate how polaron formation influences the dynamics of a two-level system (TLS) strongly interacting with a super-Ohmic environment. The last section is devoted to conclusion and outlook.

\section{\label{sec:methods}Models, Bath Representations and Methods}

The spin-boson model \cite{breuer_petruccione, Leggett} is the paradigmatic model to study from a microscopic point of view dissipative quantum dynamics.
Its Hamiltonian, $\hat H = \hat H_S + \hat H_E + \hat H_I$, describes the time evolution of a TLS $S$ interacting linearly with an environment $E$, which consists of a continuum of bosonic modes ($\hbar=1$):
\begin{eqnarray}\label{eq:hamiltonian}
 &&\hat H_S = \frac{\epsilon \hat \sigma_z}{2}, \quad \hat H_E = \int_{0}^\infty d \omega \omega \hat b_\omega^\dagger \hat b_\omega, \\ 
 &&\hat H_{I} =  \frac{ \hat \sigma_x}{2}  \int_{0}^\infty d\omega \sqrt{J(\omega)}(\hat b_\omega^\dagger + \hat b_\omega). \nonumber 
\end{eqnarray}
The bosonic operators $\hat b_\omega$, $\hat b^\dagger_\omega$ denote respectively the annihilation and creation operator of a mode of frequency $\omega$. They obey to the canonical commutation relations. The system's Hamiltonian $\hat H_S$ is defined on a two-dimensional Hilbert space, and does not commute with the interaction Hamiltonian.
The spectral density function $J(\omega)$ defines the density of the modes and the strength of the coupling of the system to each mode \cite{breuer_vacchini}.
Here, we focus on the commonly encountered power-law spectral functions of the type:
\begin{equation}\label{SDF}
    J(\omega) = 2\alpha \omega \bigg ( \frac{\omega}{\omega_c} \bigg )^{s-1} \theta(\omega-\omega_c),
\end{equation}
where $\alpha$ is a dimensionless quantity that gives a measure of the coupling strength of the system to the bath modes. The Heaviside function ensures that $J(\omega)$ has finite support in $[0, \omega_c]$. Spectral functions of this form describe a wide variety of environments in chemical, condensed phase and photonic systems \cite{WeissUlrich}. The parameter $s$ defines three classes of functions: for $s<1$ the \textit{sub-Ohmic}, for $s=1$ the \textit{Ohmic}, and for $s>1$ the \textit{super-Ohmic} spectral densities.

\begin{figure}
\includegraphics[width=0.48\textwidth]{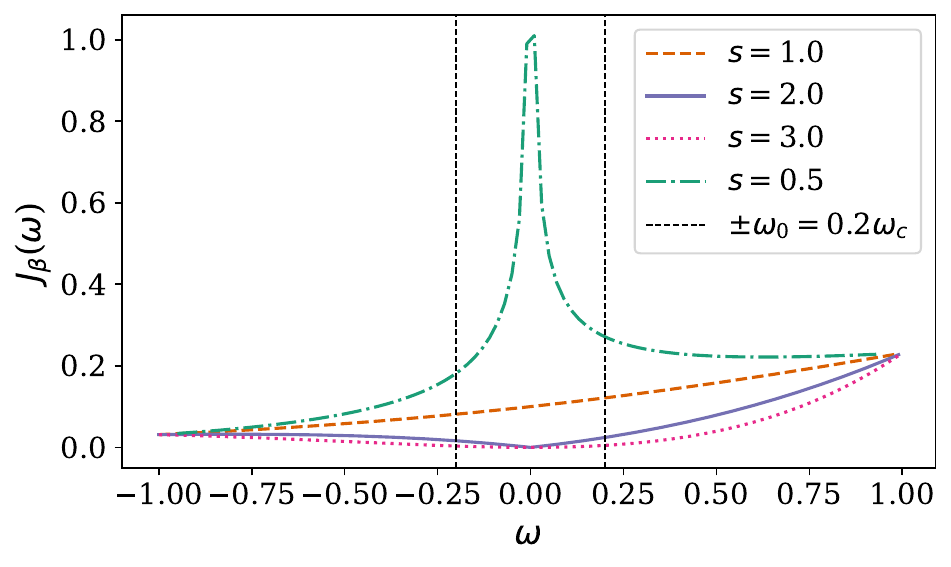}
\caption{\label{fig:therm_spectral_density} The thermalized spectral density function $J_\beta(\omega)$, at $\beta=2.0$ for different values of the degree $s$. In the super-Ohmic case ($s=2$) the spectral density is equal to zero at the origin, leading to a suppression of the low-frequency noise responsible for pure dephasing processes; such source of decoherence is instead present both in the Ohmic ($s=1$) and sub-Ohmic case ($s=1/2$) where the spectral density either does not vanish or diverges at the origin.}
\end{figure}

In what follows we restrict our attention to factorized initial states of the form $\hat \rho_{SE}(0) = \hat \rho_S(0) \otimes \hat \rho_E^\beta(0)$ where $\hat \rho_S(0) = \ket{\psi_S(0)}\hspace{-3pt} \bra{\psi_S(0)}$ is a pure state and $\hat \rho_E^\beta(0)$ is a thermal state at inverse temperature $\beta = (k_B T)^{-1}$: 
\begin{equation} \label{eq:thermalState}
    \hat \rho_E^\beta(0)=\bigotimes_\omega \frac{e^{-\beta \omega \hat b_\omega^\dagger \hat b_\omega}}{\Tr_E[e^{-\beta \omega \hat b_\omega^\dagger \hat b_\omega}]} = \bigotimes_\omega \hat \rho_\omega(\beta).  
\end{equation}
Furthermore, we assume the dynamics of system and environment to be unitary, so that the dynamics of the system alone is recovered as: 
\begin{equation}
    \hat \rho_S(t) = \Tr_E\big\{\hat U(t)\hat \rho_S(0)\otimes \hat \rho_E^\beta(0) \hat U^\dagger(t)\big\},
\end{equation}
where $U(t)=e^{-i \hat{H} t}$. The bi-linearity of the interaction operator $\hat H_I$ of Eq.~\ref{eq:hamiltonian} and the Gaussian character of  $\rho_E^\beta(0)$ guarantee that reduced state $\rho_S(t)$ of the system at time $t$ is completely determined by the environment's two time correlation function \cite{FEYNMAN1963118}:
\begin{eqnarray}\label{correlation_func_base}
    \hat S(t) &&= \int_0^\infty d \omega \Tr_E \big[\hat \rho_\omega(\beta)\hat O_\omega (t) \hat O_\omega (0)\big] \\
    &&= \int_0^\infty d\omega J(\omega)\big[e^{-i\omega t}(1 + \hat n_\omega(\beta)) + e^{i\omega t} \hat n_\omega (\beta) \big] \nonumber,
\end{eqnarray}
where the interaction operator is time evolved in the interaction picture:
\begin{equation}
    \hat O_\omega(t) = \sqrt{J(\omega)}(e^{-i\omega t}\hat b_\omega^\dagger(0) + e^{i\omega t}\hat b_{\omega}(0)),
\end{equation}
and $\hat n_\omega (\beta)$ is the Bose-Einstein occupation number at frequency $\omega$ and inverse temperature $\beta$:
\begin{equation}
\hat n_\omega(\beta) = \Tr_E \big[ \hat \rho_\omega(\beta) \hat b_\omega^\dagger \hat b_\omega \big] = \frac{1}{e^{\beta \omega}-1}.    
\end{equation}
As shown in Ref.~\cite{Tamascelli_ttedopa}, it is possible to replace the finite-temperature bosonic bath $E$ by another bath, with support on an \textit{extended} range of frequencies. The extended bath is characterized by a new spectral density function $J_\beta(\omega)$ such that its pure vacuum state correlation function matches the  thermal state correlation function (Eq. \ref{correlation_func_base}) exactly. We refer the reader to Ref.~\cite{Tamascelli_ttedopa} for full detail on the derivation; here we limit ourselves to mention two key points of the construction proposed by Tamascelli {\it et al.}. Firstly, negative frequency modes are added to the environment, by dilating the range of frequencies of the spectral density function's domain. Secondly, the temperature dependence is moved from the thermal distribution of the statistical ensemble $\hat \rho_E^\beta(0)$ to the spectral density function:
\begin{eqnarray}\label{eq:sdf_explicit}
 J_\beta(\omega) &&= \frac{1}{2}\text{sign}(\omega)J(\abs{\omega}) \left [ 1 + \coth\left ( \frac{\beta \omega}{2} \right ) \right ].
\end{eqnarray}
A thermally weighted, extended spectral density function with support on the whole real axis is thus defined. The crucial consequence is that $S(t)$ of the original thermal environment is obtained from the factorized vacuum state of the positive and negative frequency modes which make up the extended environment, i.e.

\begin{eqnarray}\label{correlation_func}
    \hat S(t) &&= \int_\infty ^\infty d \omega \bra{\text{vac}}\hat O_\omega (t) \hat O_\omega (0)\ket{\text{vac}} \\
    &&= \int_0^\infty d\omega J(\omega)\big[e^{-i\omega t}(1 + \hat n_\omega(\beta)) + e^{i\omega t} \hat n_\omega (\beta) \big] \nonumber,
\end{eqnarray}
\begin{equation}\label{vacuum}
    \ket{\text{vac}} = \bigotimes_\omega \ket{0}_\omega, \quad \hat b_\omega \ket{0}_\omega = 0 \quad \forall \omega \in \mathbb{R},
\end{equation}
and the system-extended bath interaction Hamiltonian reads:
\begin{equation}
    \hat H_I^{\beta} = \frac{\epsilon \hat \sigma_x}{2}  \int_{-\infty}^\infty d \omega \sqrt{J_\beta(\omega)}(\hat b_\omega^\dagger + \hat b_\omega).
\end{equation}
The equivalence result provided by Tamascelli \textit{et al.} in \cite{Tamascelli_ttedopa} ensures that the system's reduced dynamics determined by the interaction with the original bath, described by the spectral density $J(\omega)$ and starting from a thermal state at inverse temperature $\beta$, is equivalent to reduced dynamics determined by the interaction of the system with the extended bath with spectral density $J_\beta(\omega)$ and starting from the (pure) vacuum state. Examples of thermalized spectral densities are provided in Fig.~\ref{fig:therm_spectral_density}.  T-TEDOPA therefore shifts  thermal contributions from the initial state of the environment to the interaction strength with the harmonic oscillators of the extended environment. Interestingly enough, detailed balance condition instead of being encoded in the statistics of the initial thermal state of the oscillators $\hat \rho_\omega(\beta)$ (see Eq.~\ref{eq:thermalState}) of the original bath is now encoded in the ratio between the thermalized spectral density evaluated at opposite frequencies $\pm \omega$, namely
\begin{equation}\label{eq:balan}
    \frac{J_\beta(+\omega)}{J_\beta(-\omega)} =  e^{\beta \omega}.
\end{equation}
As we will see in the following sections, this property has interesting consequences and is fundamentally related to the determination of the heating cycle.

The second step in our simulation procedure is to map the spin-boson Hamiltonian \eref{eq:hamiltonian}    to a chain-like configuration, with only local interactions. In fact, the interaction Hamiltonian $\hat H_I$  couples the system to each mode in the bath, creating correlations between the system and all of the environmental modes at the same time. Furthermore, the coupling structure of $\hat H_I$ is unfit to be represented as an MPO, which has a linear structure, without introducing undesirable long-range interactions.  One way to overcome this problem is to exploit the observation, proven in Ref.~\cite{chain_mapping_original}, which establishes that a system linearly coupled to a reservoir with a spectral density $J(\omega)$ can be unitaliry transformed into a semi-infinite chain where the system is coupled only to the first mode of a new chain-like environment. By mapping the interaction onto nearest-neighbor couplings, this approach enables simulations to compute the complete many-body dynamics of both the system and the environment \cite{WoodsChin}. Once the chain representation is obtained, non-perturbative simulations of the open quantum system can be conducted using tensor network techniques, specifically Thermalized Time Evolving Density operator with Orthogonal Polynomials (T-TEDOPA) \cite{Tamascelli_ttedopa}.

The chain mapping transformation starts from the introduction of the operators
\begin{equation}\label{eq:chain_modes}
    \hat{c}_n^\dagger= \int_{-\infty}^\infty d\omega U_n(\omega) \hat b_\omega^\dagger, \quad U_n(\omega) = \sqrt{J_\beta(\omega)} \Tilde{p}_n(\omega),
\end{equation}
with  $\Tilde{p}_n(\omega)$ indicating polynomials that are orthogonal with respect to the measure $d\mu= J_\beta(\omega)d\omega$. In terms of the bosonic operators  $\hat c_\omega$, $\hat c^\dagger_\omega$, satisfying the canonical commutation relations, the Hamiltonian \eref{eq:hamiltonian} reads
\begin{align} \label{eq:chainHam}
\hat{H}^C &= \hat{H}_S + \hat{H}_I^C + \hat{H}_E^C  \\
& = \hat{H}_S + g_0 \hat \sigma_x (\hat{c}_0+\hat{c}_0^\dagger)+ \nonumber \\
& \sum_{k=0}^\infty \omega_j \hat c_k^\dagger \hat c_k + \sum_{k=1}^\infty g_k (\hat c_k^\dagger \hat{c}_{k-1} + \text{H.c.}). \nonumber
\end{align}
We refer the reader to Refs.~\cite{chain_mapping_original,Tamascelli_ttedopa} for a full account on the (T-)\-TEDOPA chain mapping. Here we limit ourselves to mention that the chain modes frequencies $\omega_k$ and coupling constants $g_k$ depend on the orthogonal polynomials $\Tilde{p}_n(\omega)$ which, in turn, depend on the spectral density $J_\beta(\omega)$. While in some cases orthogonal polynomials for specific measures $d\mu$ can be analytically found,  stable numerical routines, such as ORTHPOL~\cite{orthpol}, are in general exploited to determine the chain coefficients.~\footnote{The numerical methods employed use the software packages available at: \url{https://github.com/tfmlaX/Chaincoeffs}}.

In conclusion, the T-TEDOPA chain mapping reformulates the OQS problem as a $1D$ many-body problem with only nearest-neighbor interactions: the dynamics can be efficiently simulated with tensor network methods, and in particular with the recently developed bond-adaptive one-site time-dependent variational principle algorithm for MPS time evolution, which has been used to obtain the numerical results presented in this paper (see Appendix \ref{app:mps}) \cite{Dunnett_Chin_2021_evolving}.

\section{\label{sec:results}Setup}

The features of the environment, and therefore of the noise it induces on the system, are dictated by the spectral density function and by the initial state of the environment which, in our setting, are synthesized by Eq.~\eref{eq:sdf_explicit}. The behavior of the spectral density around the origin plays an important role: in the sub-Ohmic case $s=1/2$, the thermalized spectral density $J_{\beta}(\omega)$ diverges at zero frequency; in the Ohmic case $s=1$ it has a finite value $J_{\beta}(0)>0$ whereas it is equal to zero in the super-Ohmic case $s \ge 2$. Therefore we expect pure dephasing processes, happening at low frequency and typically associated to inelastic scattering processes, to account for a consistent part of the system-environment exchanges in the Ohmic and sub-Ohmic cases. Since there is no energy transfer, pure dephasing processes do not affect the populations of the TLS. Conversely, a suppression of the pure dephasing noise should occur for the super-Ohmic environment. The interaction with modes of frequency around the energy gap of the TLS $\omega \approx \epsilon$, where $J_{\beta}(\omega)$ is non-zero for any value of $s$, can drive a transition through emission/absorption processes. Finally, non-resonant high frequency modes cause the formation of polaronic states of the TLS and of the environment, leading to the renormalization of the frequency gap of the dressed TLS \cite{Haroche:993568}.

We remark that in what follows we take full advantage of the possibility provided by T-TEDOPA to inspect not only the open quantum system's degrees of freedom but also those of the surrounding environment. Indeed, measurement performed on the harmonic oscillators of the chain mapped environment, can be used to determine properties of the oscillators in the original ``star configuration''. This provides a most powerful tool to underpin the fundamental mechanisms determining the evolution of the open quantum system. 

Before proceeding with the presentation of our results, we provide some detail on the setting used throughout this work.
We consider spectral densities of the form \eref{SDF}, namely Ohmic spectral densities with hard-cutoff at $\omega_c$.
The TLS energy gap is set to $\epsilon=0.2\omega_c$. The open quantum system dynamics are explored in the low and  high temperature regimes defined by the parameter $\kappa=\epsilon \beta$ for the choice $\kappa = 400$ and $\kappa = 0.4$ respectively.
For the sake of definiteness, and without loss of generality, in what follows we set $\omega_c=1$. In order to look at the impact of the shape, i.e. Ohmic, sub-Ohmic, etc., of the spectral function on the underlying physics of the T-TEDOPA, we use a $s$-dependent overall system bath coupling constant $\alpha$ (see Eq.~\eref{SDF}), i.e. 
\begin{equation} \label{eq:coupFact}
\alpha \equiv \alpha(s)= \alpha'/\epsilon^s,   
\end{equation}
with $\alpha'= 0.01$. This choice makes the Markovian TLS decay rate independent of $s$, enabling easier comparison of the results (see Appendix \ref{app:alpha} for more details). 

In order to enable efficient simulation of the evolution of the system and of the environment degrees of freedom we  exploit the T-TEDOPA chain mapping described in the previous section. As to allow for numerical simulation we will truncate the resulting semi-infinite chain of harmonic oscillators after $N=120$ sites: this choice allows to avoid any finite-size artifacts within the considered $\omega_c t = 100$ simulation time. For more details on the parametrization of the  MPS/MPO used for the DTDVP time evolution we refer the reader to Appendix~\ref{app:mps}.

\section{\label{sec:pure_dephasing} Evidence of pure dephasing processes in the sub-Ohmic environment}

\begin{figure}
\includegraphics[width=0.43\textwidth]{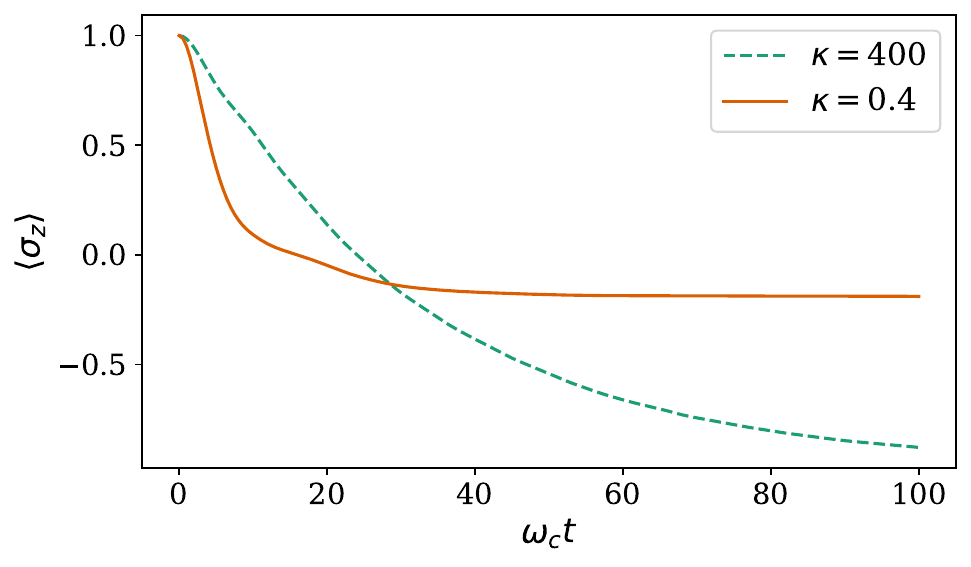}

\caption{\label{fig:spin_s=0.5} Sub-Ohmic case $s=1/2$. The TLS population $\langle \sigma_z\rangle$ as a function of time, for the initial state of the system $\ket{\psi_S(t=0)} = \ket{1}$, at low  ($\kappa=400$) and high ($\kappa=0.4$) temperature .}
\end{figure}

\begin{figure}
\includegraphics[width=0.48\textwidth]{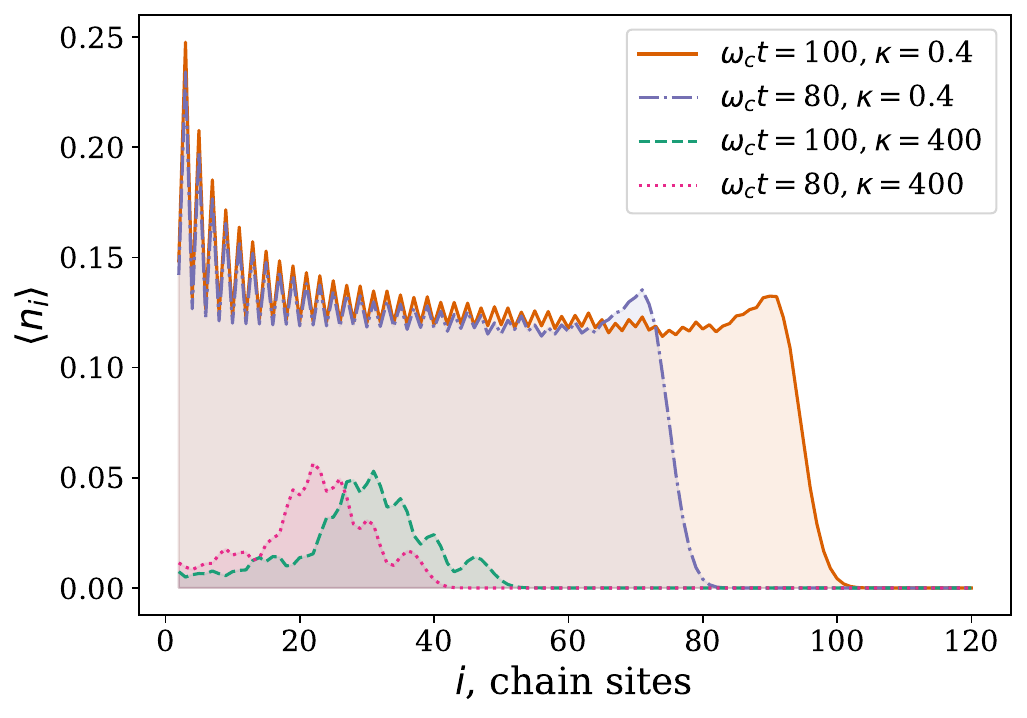}
\caption{\label{fig:chain_s=0.5} Sub-Ohmic case $s=1/2$. The average occupation number $\langle n_i \rangle$ of the chain modes as function of  the chain site index $i$ in the low ($\kappa=400$) and high ($\kappa=0.4$) temperature regime at the times $\omega_c t=80$ and $\omega_c t=100$.
}
\end{figure}

We start by considering the TLS, initially set in the excited $\ket{\psi_S(t=0)} = \ket{1}$ state, interacting with a sub-Ohmic environment ($s=1/2$) at low- and high-temperature. As mentioned in the previous section, in T-TEDOPA the oscillators in the chain are initially in the vacuum state~\eref{vacuum}. 
The time evolution of the $\langle \hat \sigma_z \rangle $ is shown in Fig. \ref{fig:spin_s=0.5}: {in the high temperature case, the system quickly converges to a steady state that is nearly, but not exactly, zero;} at low temperature, instead, the value of $\langle \hat \sigma_z \rangle $ approaches a markedly negative value: the system is relaxing toward its ground state. 
Figure~\ref{fig:chain_s=0.5} shows the average occupation number of the chain modes. At low temperature, a finite number of excitations is introduced in the chain by the system, forming a \textit{wave packet} that propagates along the chain; at high temperature, on the other hand, the system never stops introducing new excitations on the chain, creating a steady wave which propagates faster than in the low temperature case.  
We moreover observe that the creation of new excitations in the chain persists well after the system has reached the steady state. Considered that the chain Hamiltonian terms in the last line of equation \eref{eq:hamiltonian} conserve the number of excitations, this behavior witnesses the onset of a non-equilibrim steady state, where energy is continuously exchanged between the system and the chain.

Figure~\ref{fig:extended_env_s=0.5} shows the average occupations in the extended bath (see Appendix \ref{app:grid} for details on the frequency sampling), obtained by inverting the chain mapping transformation. The conservation of the total number of excitations in the chain mapping ensures that the TLS injects only a finite number of excitations into the extended environment at low temperatures as it decays to the ground state. Consequently, only modes resonant with the system transition energy are populated, leading to an emission spectrum. In contrast, in the high-temperature regime, the continuously generated modes on the chain give rise to an unbounded growth of the extended bath modes around the origin, qualitatively following the shape of the thermal effective spectral density shown in Fig. \ref{fig:therm_spectral_density}. Indeed, as the total excitation energy of the chain and extended environment must be the same, we can understand the continuous growth of the peaks in the extended picture as resulting from the growing area under the chain population curves in Fig. \ref{fig:chain_s=0.5}, due to the persistent tail that is left behind as the leading wave front moves along the chain. {We note that while the number of excitations at each individual chain site remains bounded, the total number of excitations across the entire chain continues to grow indefinitely. Therefore, simulating extended time periods requires longer chains and, consequently, additional computational resources.}

We note that the population of low frequency bath modes in the extended picture is linked to pure dephasing noise which is the dominating process in the high temperature regime for Ohmic and sub-Ohmic environments \cite{Leggett}. Previous evidence of pure dephasing noise in the Ohmic ($s=1$) environment has been discussed in Ref.~\cite{Dunnett_Chin_2021}, but the broad Gaussian pure dephasing peak masks all other thermal processes, such as energy exchange. Here, by contrast, the sharp peak around zero frequency  enables us to resolve the two small peaks around  $\omega\approx\pm \epsilon$, which correspond to emission/absorption processes of the TLS into the extended environment.

\begin{figure}
\includegraphics[width=0.48\textwidth]{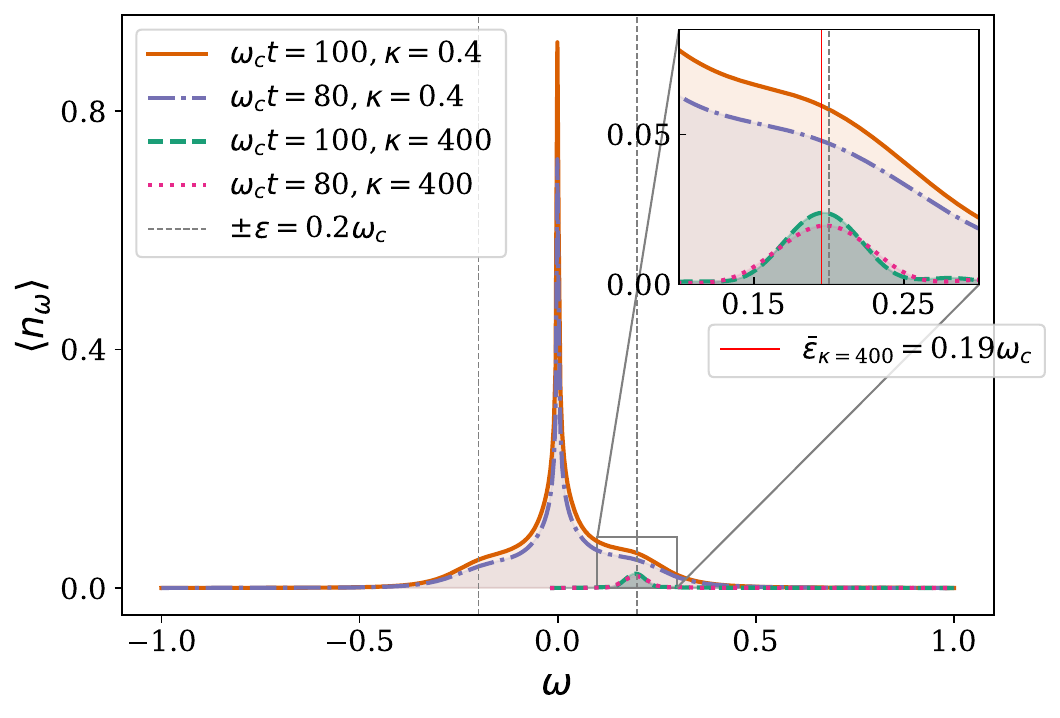}
\caption{\label{fig:extended_env_s=0.5} Sub-Ohmic case $s=1/2$. The average occupation number $\langle n_\omega \rangle$ of the extended environment modes as a function of $\omega$, at times $\omega_c t =80$ and $\omega_c t =100$  and different  temperatures.  At low temperature ($\kappa=400$), a single small peak centered around the TLS transition frequency $\epsilon = 0.2 \omega_c$ is clearly visible in the inset. At high temperature ($\kappa=0.4$), the spike at zero frequency can be associated with pure dephasing noise, whereas the small peaks at $\pm \epsilon$ can be  related to energy exchange processes between the system and the environment.}
\end{figure}
 
\section{\label{sec:cycle} Testing the thermal cycle hypothesis in the super-Ohmic environment}
In this section we discuss the super-Ohmic scenario ($s=2$). As mentioned before, in the presence of super-Ohmic environments, low frequency noise is suppressed and thermal processes dominate the exchanges of energy between the system and the extended environment. Super-Ohmic environments thus provide the ideal setting to test the thermal cycle hypothesis which, we remind, should manifest as a continuous growth of the average occupation number of the environmental modes having frequency close to the transition energy of the system.

\begin{figure}
\includegraphics[width=0.43\textwidth]{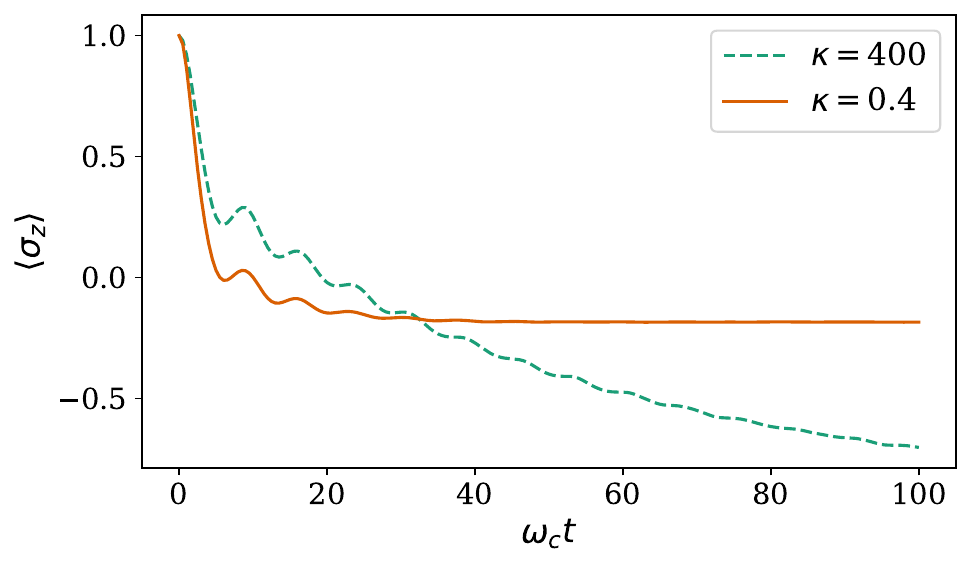}
\caption{\label{fig:spin_s=2} Super-Ohmic case $s=2$. The TLS average population $\langle \sigma_z \rangle$ as a function of time for $\ket{\psi_S(t=0)} = \ket{1}$, at low ($\kappa = 400$) and high  ($\kappa=0.4$) temperature. }
\end{figure}

\begin{figure}
    \centering
    \includegraphics[width=0.48\textwidth]{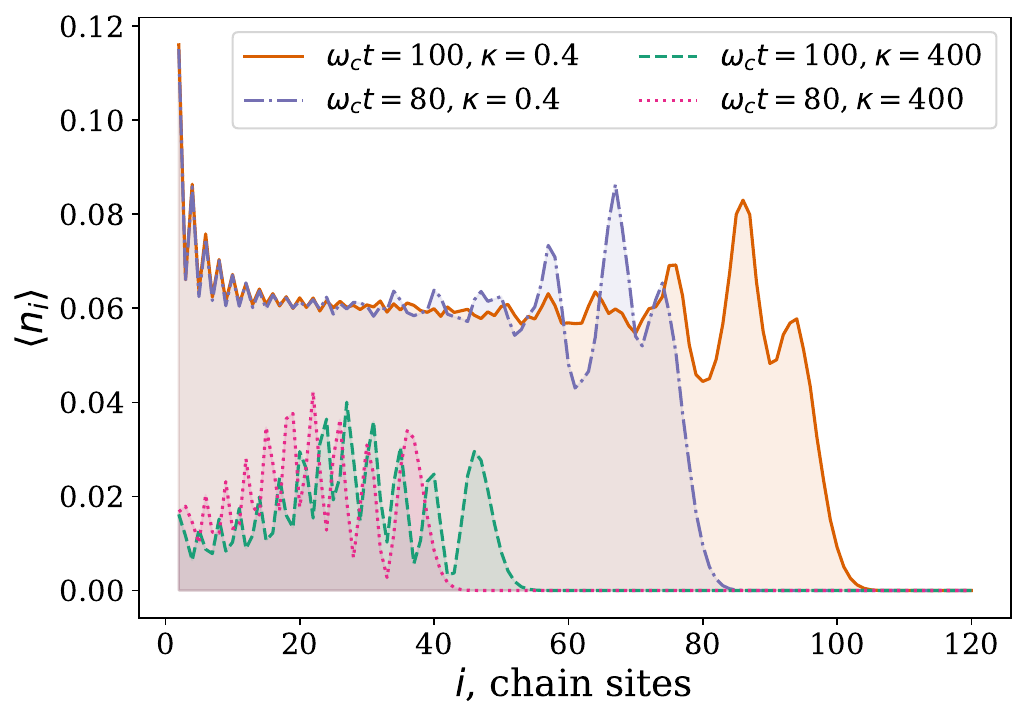}
    \caption{\label{fig:chain_s=2} Super-Ohmic case $s=2$. The average occupation number $\langle n_i \rangle$ of the chain modes as function of the chain site index $i$ at the times $\omega_c t=80$ and $\omega_c t=100$.}
\end{figure}

As Fig. \ref{fig:spin_s=2} shows, the time evolution of the system observable $\langle \hat \sigma_z \rangle$ in the low and high temperature regime is quite similar to the one shown in Fig.~\ref{fig:spin_s=0.5} (referring to the sub-Ohmic case). 
Indeed, at short times it is possible to discern small oscillations suggesting some form of coherent dynamics. The evolution of the average occupation of the chain oscillators for the super-Ohmic case considered here, shown in Fig.~ \ref{fig:chain_s=2}, shares some features with the sub-Ohmic case (Fig.~\ref{fig:chain_s=0.5}): at high temperature a continuous stream of excitations is continuously being injected by the TLS into the chain, whereas at low temperature  the wave of excitations has a pulse-like shape, corresponding to a a finite total number of excitations travelling along the chain. On the other side, we can observe that  average occupation of the chain sites is much smaller  than in the sub-Ohmic case: the rate at which excitations are ``pumped'' into chain by the interaction with the system is smaller in the super-Ohmic scenario.
\begin{figure}
\includegraphics[width=0.48\textwidth]{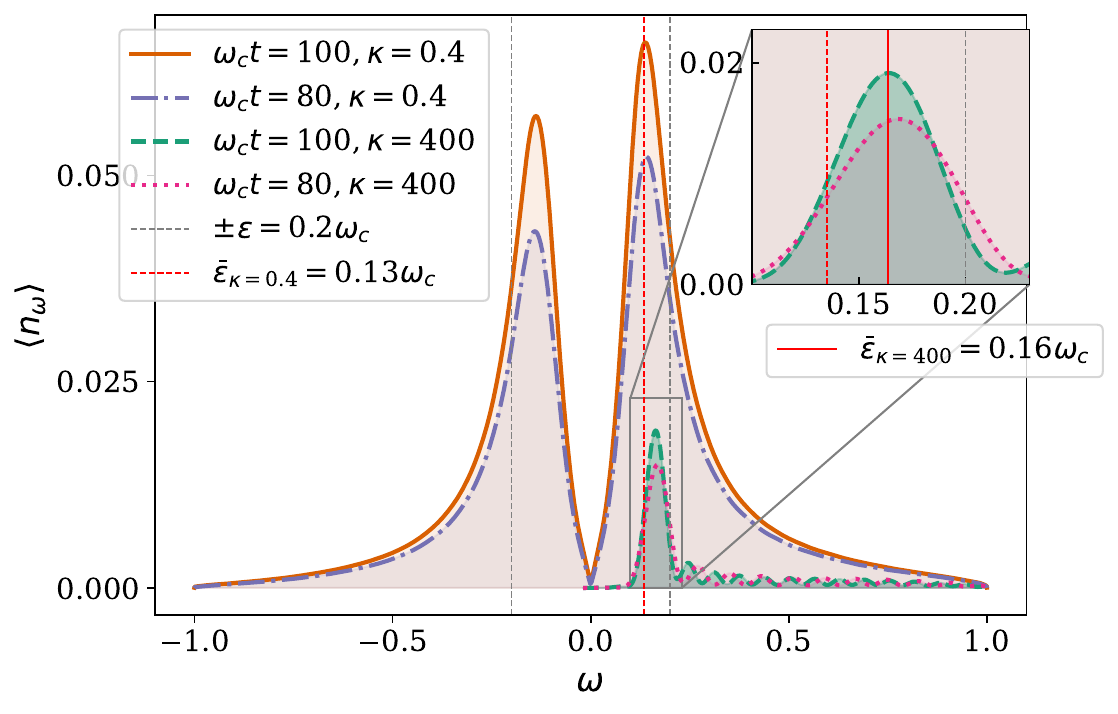}
\caption{\label{fig:extended_env_s=2} Super-Ohmic ($s=2$) case. The average occupation number of the extended environment modes as a function of $\omega$ at low/high ($\kappa = 400/0.4$) temperatures determined at  times  $\omega_c t =80$ and $\omega_c t =100$. At low temperature ($\kappa=400$), a small peak around the renormalized  transition energy of the system, i.e. $\omega \approx +\bar{\epsilon}_{\kappa=400}$; at finite temperature ($\kappa=0.4$), two peaks appear at $\omega \approx \pm \bar{\epsilon}_{\kappa=0.4}$, corresponding to absorption ($- \bar{\epsilon}_{\kappa=0.4}$) and emission ($+ \bar{\epsilon}_{\kappa=0.4}$) of energy from the TLS into the extended environment.}
\end{figure}
Fig.~\ref{fig:extended_env_s=2} provides instead a conclusive evidence of the heating cycle hypothesis. As a matter of fact, the suppression of low frequency noise allows for the formation of two well visible peaks close to the (renormalized, vide infra) system transition energy $\omega \approx \bar{\epsilon}_{\kappa=0.4}<\epsilon$. The height of such peaks, moreover, goes on increasing even when the system has reached a stationary state, thus confirming the continuous creation of pairs of excitations of opposite frequency  as predicted by the thermal cycle hypothesis. It is furthermore easy to check that the population unbalance between opposite frequency modes satisfies, at any time, the detailed balance condition \eref{eq:balan}.

Beside supporting the heating cycle hypothesis, these observations well clarify that when the thermal equilibrium state is reached the energy goes on flowing from the system to the extended environment and vice-versa, leading to the cyclic behavior~ \cite{Dunnett_Chin_2021}:  the overall system-environment stationary state is therefore a stationary non-equilibrium state.  
We also observe that the system-bath interaction creates entanglement between the system and the bath degrees of freedom. The new (polaronic) eigenstates of system and environment are no longer system/bath factorized and the renormalized transition energy of the system $\bar{\epsilon}_{\kappa=0.4}$ is red-shifted ($\bar{\epsilon}_{\kappa=0.4}<\epsilon$). 
The value of the renormalized energy, moreover, can be estimated by looking at the frequency values corresponding to the maxima of the peaks in Fig.~\ref{fig:extended_env_s=2}. We remark that the renormalized energy gap is temperature dependent, and we notice that $\bar{\epsilon}_{\kappa=0.4}< \bar \epsilon_{\kappa=400}$.

\begin{figure}
\includegraphics[width=0.48\textwidth]{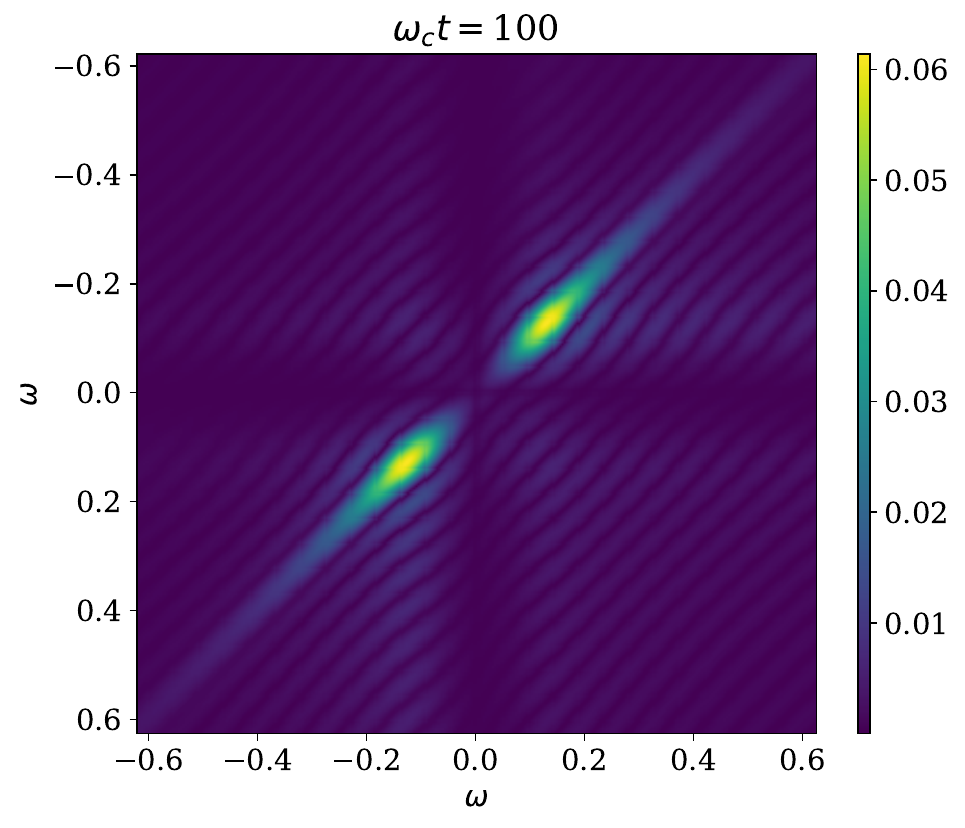}
\caption{\label{fig:correlations} High temperature ($\kappa=0.4$), super-Ohmic ($s=2$) case. The correlation function $C(\omega, \omega')$ at the final time-step of the simulation. Modes of opposite frequency are strongly correlated, evidencing the heating cycle. These anti-correlated peaks vanish as temperature is reduced.}
\end{figure}

As a final proof of the thermal cycle, we examine the correlations between environmental modes of opposite frequency value at finite temperature in the extended environment. As discussed before, pairs of excitations having opposite frequencies are expected to be highly correlated around the energy of the TLS gap, since emission/absorption processes must happen at rates determined by the relation \eref{eq:balan}. Fig.~\ref{fig:correlations}, showing the frequency-frequency correlations
\begin{equation}
    C(\omega, \omega') = \langle b_\omega^\dagger b_{\omega'}^\dagger \rangle - \langle b_\omega^\dagger\rangle \langle b_{\omega'}^\dagger \rangle
\end{equation}
for the high temperature super-Ohmic case, confirms the presence of these predicted antidiagonal correlations. That this arises from thermal effects is also evidenced by the continuous vanishing of these peaks, as the temperature is lowered (not shown).  We remark that a similar correlation landscape is expected also in other thermalization schemes, as the thermofield approach~\cite{devegaban}, where an auxiliary environment of bosonic modes of negative frequencies is entangled via two-mode squeezing with a positive frequency bath in order to purify the  mixed thermal state of the physical environment. 
\begin{figure}
\includegraphics[width=0.48\textwidth]{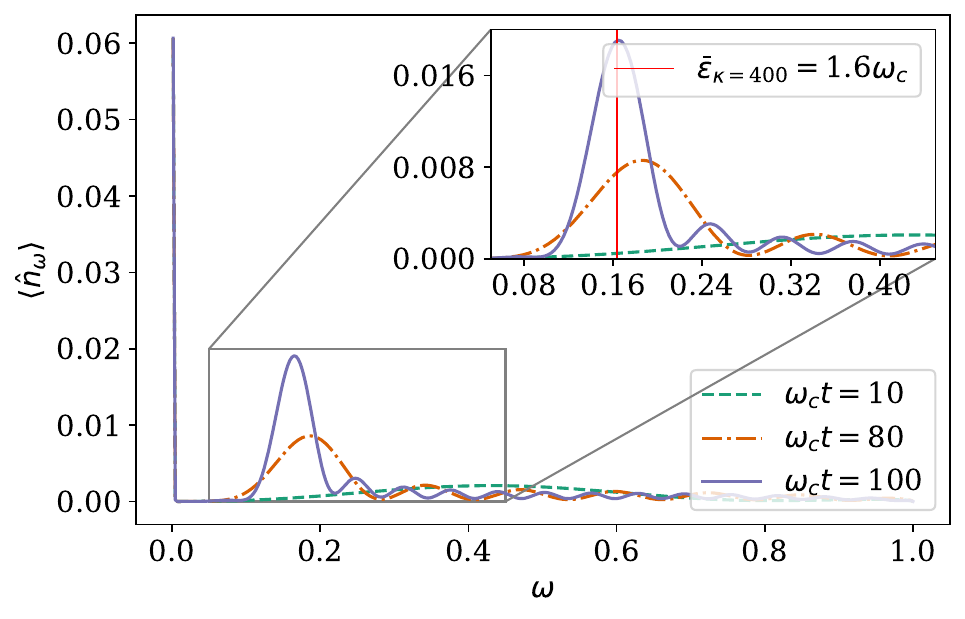}
\caption{\label{fig:back_low} Average occupation number $\langle \hat{n}_\omega\rangle$ of the physical (positive frequency) environmental modes of the super-Ohmic ($s=2$) environment at low temperature ($\kappa=400$) at different times. A peak due to the emission of energy from the TLS to the bath is clearly visible at $\omega \approx \bar{\epsilon}_{\kappa=400}$.}
\end{figure}

\begin{figure}
\includegraphics[width=0.48\textwidth]{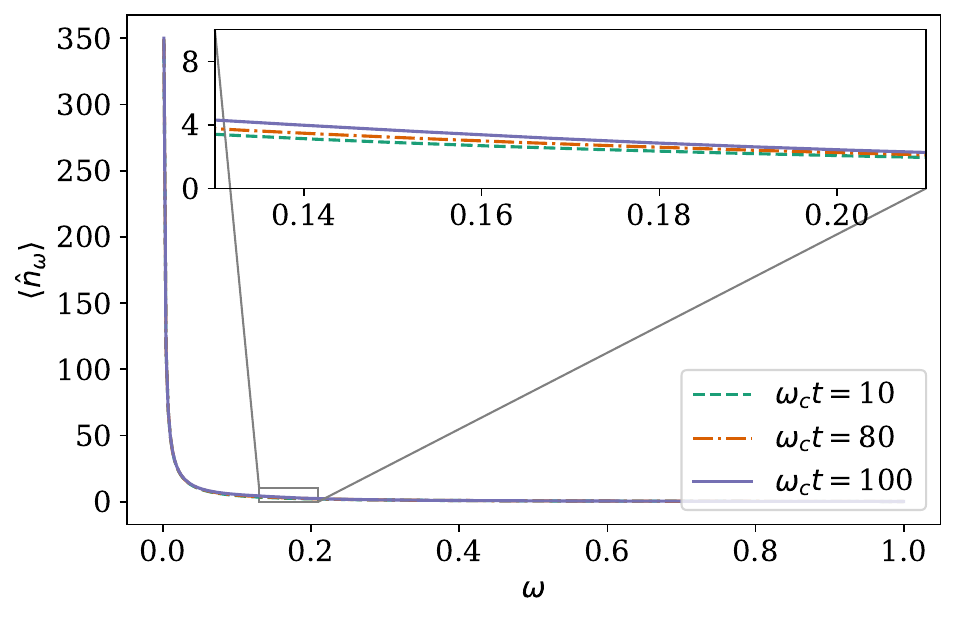}
\caption{\label{fig:back_high} Average occupation number $\langle \hat{n}_\omega\rangle$ of the physical (positive frequency) environmental modes of the super-Ohmic ($s=2$) environment at high  temperature ($\kappa=0.4$) at different times. The environment can be approximated by a heat bath.}
\end{figure}

We conclude this section by discussing the evolution  of the original bath, consisting of positive frequency ``physical'' harmonic oscillators, determined by the interaction with the system. So far, in fact, we have presented results referring to thermalized extended environments by looking at both the evolution of the chain oscillators and of the chain normal modes. Recovering the value of the observables of the physical modes is therefore an important step to complete the description presented in this work. We can do so by reverting the T-TEDOPA transformation 
described in Section~\ref{sec:methods}. We refer the reader to Appendix~\ref{app:thermofield} for details on such inversion procedure; here we stress that it is the first time, to the best of our knowledge, that such a back-mapping from the extended T-TEDOPA environment to the original environment is defined.
In Fig. \ref{fig:back_low} we show the average occupation number for the modes  in the original domain $[0,\omega_c]$ for the super-Ohmic case in the low temperature regime $\kappa = 400$ and the system starting from its excited state $\ket{1}$. As discussed above (see also Fig.~\ref{fig:spin_s=2}) this setting the system decays to its ground state by emitting energy at its renormalized transition frequency $\omega \approx \bar{\epsilon}_{\kappa=400}$. Such emission process perturbs significantly the state of the environment, creating excitations in the modes $\omega \approx \bar{\epsilon}_{\kappa=400}$. We remark that such persisting changes of the environmental state due to the interaction with the TLS in the environment are a potential signature of non-perturbative, or non-Markovian dynamics: the environment state can not be described as unperturbed, or memory-less \cite{breuer_vacchini, rivas2012open,lacroix2021unveiling}. 

At high temperature, the interaction with the system leads to much less evident perturbations of the state of the environment. Fig. \ref{fig:back_high} shows that the average occupation number of the bath modes only slightly deviates, at the final time $\omega_c t=100$, from its initial value determined by the Bose-Einstein density distribution $\hat n_\omega = 1/(e^{\beta \omega}-1)$. In the physical environment there is no sign of the peaks clearly visible in the extended environment of Fig.~\ref{fig:extended_env_s=2}: interestingly, it seems that the extended environment is not merely a computational tool needed to gain computational efficiency, but also makes manifest some interesting physical properties that are not visible in the original environment picture, such as the renormalized energy of the finite temperature TLS interacting with the bath.  Of course, the ability to look at the behavior of the bath in the physical picture could, in future work, provide interesting insights for non-equilibrium problems such as heat flows between environments, or questions related to quantum thermodynamics and energy management in quantum devices \cite{auffeves2022quantum}. 

In both the low and high temperature regimes, after the system's dynamics has relaxed, the total number of physical excitations reaches a stable value, as expected: the growth in the occupations happening in the extended bath is a mathematical representation of the thermal fluctuations that excite and de-excite the system, but does not reflect in an infinite growth of physical modes.

\section{A change in the ground state of the super-Ohmic environment}\label{sec:s=3}
In the previous sections we have provided evidence of the fact that T-TEDOPA, by treating the system and the environment on the same footing and thus leaving the environmental degrees of freedom available for inspection, provides a most powerful tool for the understanding of the fundamental mechanisms underpinning open quantum systems dynamics, such as the onset of dephasing and heating cycle. 

Here we exploit T-TEDOPA as to further investigate the onset of polaron formation and its impact on the dynamics of the TLS. Polaron formation is associated with the dressing of the TLS dynamics by the high frequeency modes on an environment, leading to renormalisation of TLS energy gaps, or tunneling matrix elements \cite{Leggett,Chin_QPT_sub-ohmic_SBModel_ansatz}. To this end it is expedient to consider a  super-Ohmic scenario with $s=3$: the higher degree allows, on the one side, to further suppress low frequency noise and enhance high-frequency modes and, on the other, to increase, for our choice of the coupling factor $\alpha(s)$ (see Eqs.~\eref{eq:coupFact}) the system-bath overall coupling. Moreover, as to isolate polaron formation from other effects, in this section we will restrict our attention to the low temperature regime ($\kappa=400$). 
\begin{figure}
\includegraphics[width=0.48\textwidth]{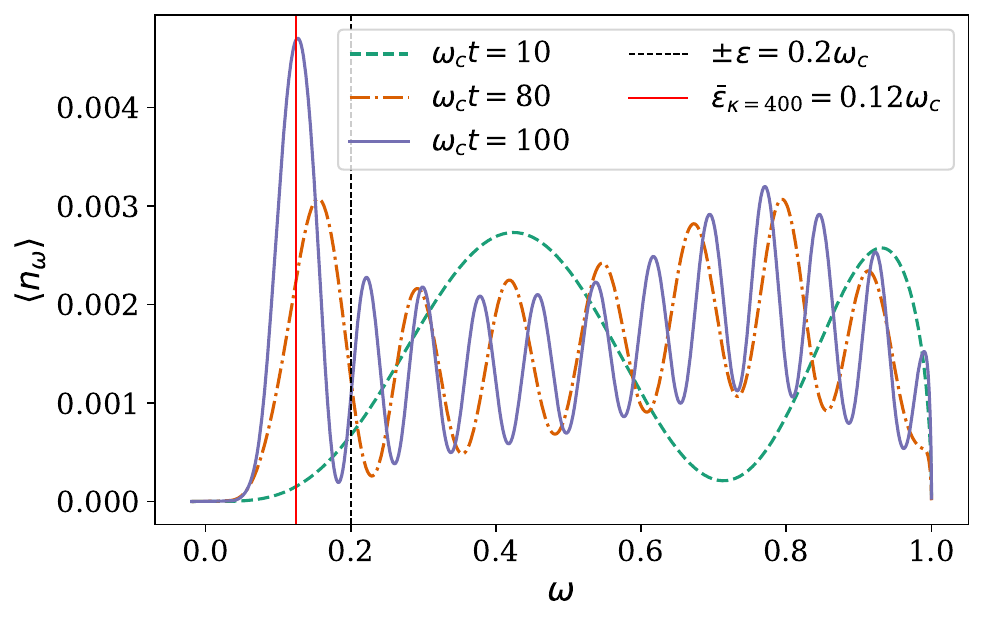}
\caption{\label{fig:back_low_s_3} Average occupation number $\langle \hat{n}_\omega\rangle$ of the extended environment modes in the super-Ohmic $s=3$ case  at low  temperature ($\kappa=0.4)$ at different times. Unlike the sub-Ohmic and $s=2$ cases, a broad range of excitations appear across all frequencies, in addition to the absorption-emission peak that appears at a strongly renormalized TLS energy gap.  }
\label{fig:s=3pops}
\end{figure}
\begin{figure}
\includegraphics[width=0.48\textwidth]{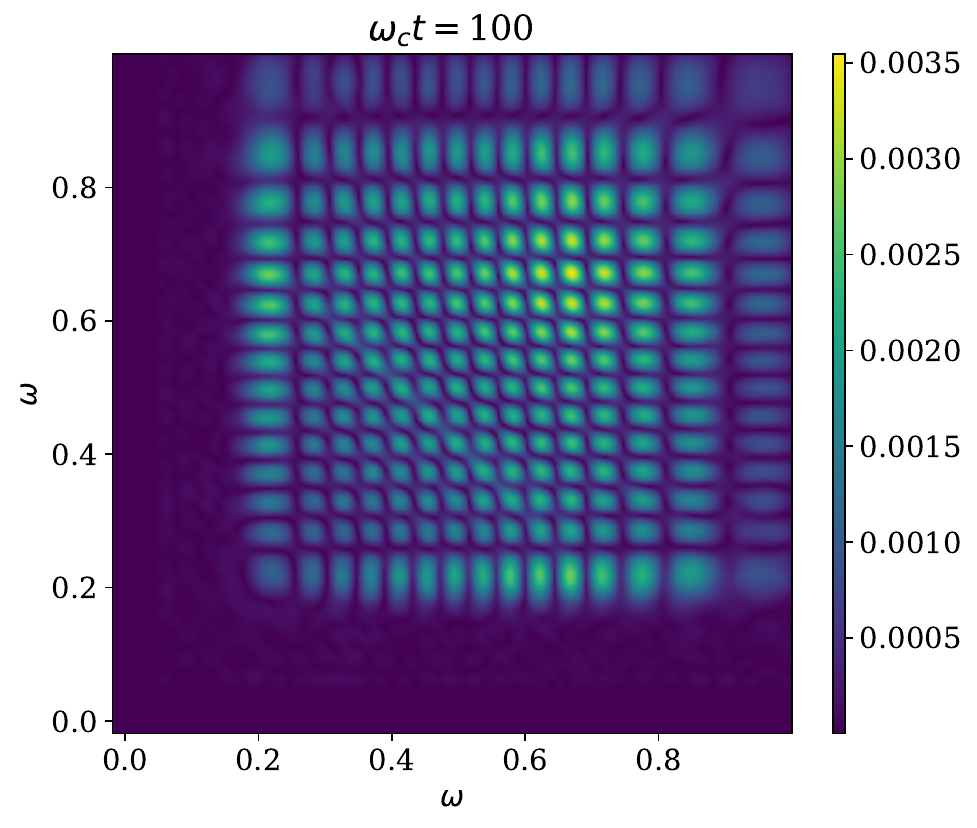}
\caption{\label{fig:correls=3} Super-Ohmic ($s=3$) environment at low temperature ($\kappa = 400$). The correlation function $C(\omega, \omega')$ at the final time-step of the simulation exhibits a grid-like pattern.}
\end{figure}
In the presence of ultra-strong coupling, the interaction of the system with the $s=3$ environment populates many environmental modes. This is clearly shown in Fig.~\ref{fig:back_low_s_3}: as time passes we see the formation of a higher peak, centered around the renormalized transition energy of the system, followed by number of smaller peaks extending across \emph{all} frequencies up to $\omega_c$. Interestingly, the frequency-frequency correlation plot shown in Fig.~\ref{fig:correls=3} reveals that, at the final time $\omega_c t = 100$, there are non-negligible correlations among all the populated modes. In further contrast to the correlation plots for the $s=2$ case, we also see that the largest intensities occur at \emph{high} frequencies, and there is negligible correlation at the (renormalized) absorption-emission peak. This pattern of correlation, we shall show, is consistent with the formation of polaron states of the TLS.      

\begin{figure}
\includegraphics[width=0.48\textwidth]{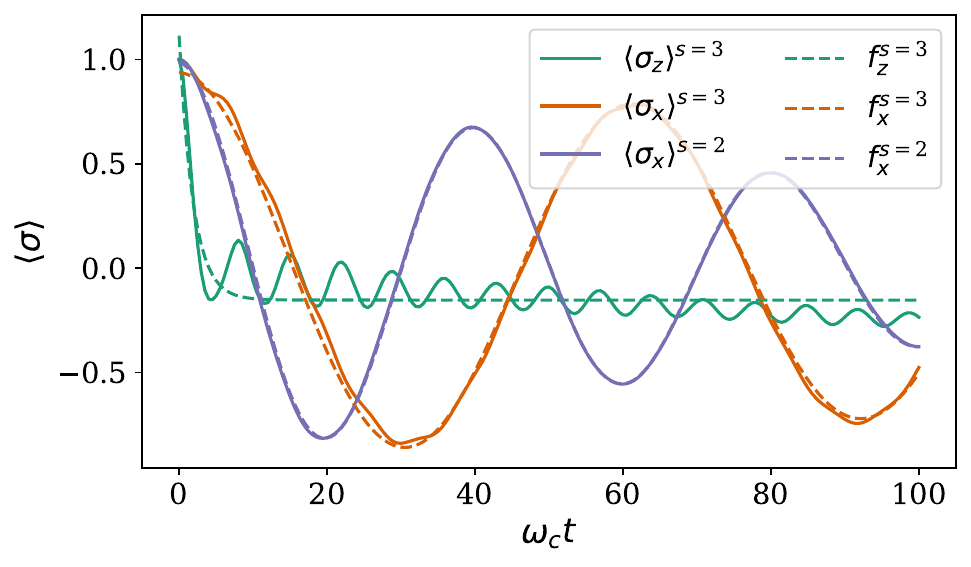}
\caption{\label{fig:spin_s=3} The population ($\langle \sigma_z \rangle$) and coherence ($\langle \sigma_x \rangle$) dynamics in the super-Ohmic $s=3$ case at low temperature ($\kappa=400$). Green solid line: $\ket{\psi_S(t=0)} = \ket{1}$; the initial decay is well fitted by the curve $f_z^{s=3}(t) = a_z e^{-\Gamma_z t}+c_z$, with $a_z=1.268$, $\Gamma_z=0.521$ and $c_z=-0.153$ (dashed green line). Solid blue and solid orange lines: the dynamics of the coherence $\langle \sigma_x \rangle$, with the system starting from $\ket{\psi_S(t=0)}= \frac{1}{2}(|1 \rangle + |0 \rangle)$ for, respectively, $s=2$ and $s=3$. Dashed blue and dashed orange lines correspond to the fitting curves for the $s=2$ and $s=3$ cases (fitting parameters provided in the main text).}
\end{figure}

The evolution of the the TLS is shown in Fig.~\ref{fig:spin_s=3}. In particular, the green solid line corresponds to the evolution of $\langle \sigma_z(t) \rangle$ when the system starts from the excited state $\ket{\psi_S(t=0)} = \ket{1}$. We can appreciate that, even if the interaction with the environment is stronger than in $s=2$ case considered in the previous section, the system does not emit all of its energy to the environment and, therefore, does not relax, within the considered time, toward the expected $\langle \sigma_z \rangle =-1$ ground state. Instead, after an initial steep decay, the $z$-polarization of the system shows a very slow decrease accompanied by oscillations that are much more pronounced than in the $s=2$ case, as if the presence of the bath were somehow inducing some sort of  ``coherent driving'' of the system. Even more interestingly, the dashed (orange) line shows that, when initialized in the state $\ket{+} = (\ket{0}+\ket{1})/\sqrt{2}$, namely the eigenstate of $\sigma_x$ belonging to the eigenvalue $+1$, the system exhibits a precession around the $z$ axis at frequency $\omega_x = 0.102 \approx \bar{\epsilon}_{\kappa=400}$;  such precession frequency  corresponds to the location of the highest peak of Fig.~\ref{fig:back_low_s_3}, which we can once more associate to the renormalized transition frequency of the system. By comparing such precession motion with the one obtained in the $s=2$ case (solid blue line of Fig.~\ref{fig:spin_s=3}) it is clear that the frequency and damping rate of the oscillations of $\langle \sigma_x(t)\rangle$ is smaller in the $s=3$ case, even though the system-bath coupling is much stronger in the latter case. More precisely, the  $\langle \sigma_x(t) \rangle$ is well fitted by the curve $f_x^{s=3}(t) = a_x \cos(\omega_x t)e^{-\Gamma_x t}$, with  $a_x=0.937$, $\omega_x=0.102$ and  $\Gamma_x=0.003$, whereas for $s=2$ the fitting curve $f_x^{s=2}(t)$ has the same form as $f_x^{s=3}$ but with coefficients  $a_x$=0.987, $\omega_x$=0.157 and  $\Gamma_x=0.010$. {We also remark the appearance of high frequency oscillations for the observable $\langle \hat \sigma_z \rangle^{s=3}$, which we attribute to ringing artifacts, introduced most likely due to the hard cut-off in the frequency domain of the spectral density function of Eq.\ref{SDF}}.

\begin{figure*}
    \centering
    \includegraphics[width=0.98\textwidth]{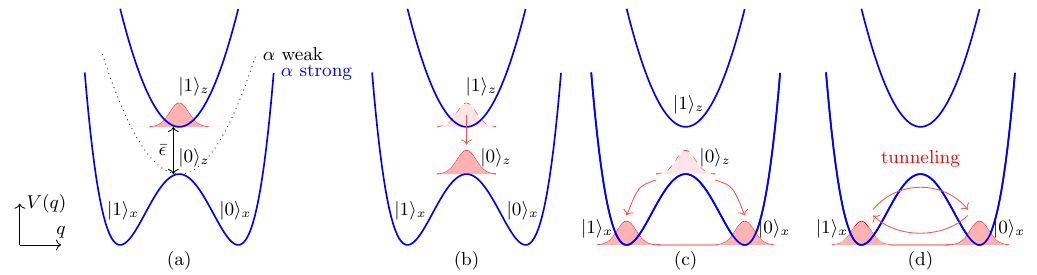}
    \caption{Polaron dynamics for strongly coupled super-Ohmic environments. (a) The strong system-environment coupling to the collective `reaction coordinate' mode creates an adiabatic double well potential for this mode, parameterised by its classical displacement $q$. Importantly, on such adiabatic surfaces, the `local' eigenstates of the TLS change continuously as a function of $q$; in the strong coupling case the lowest energy states require the TLS to be eigenstates of $\sigma_x$, as opposed to the weak coupling case (black) in which they would be eignestates of the bare TLS Hamiltonian, i.e. $\sigma_z$. (b) The TLS excitation and the initial environment wave function (Gaussian vacuum) decay onto the lower potential surface where (c) the reaction coordinate state splits and relaxes towards the two local minima. (d) due to the finite overlap of the oscillator wave functions in each potential minimum, the TLS can undergo renormalized (polaronic) tunneling between minima, leading to slow oscillations in $\langle \sigma_x (t) \rangle$ which, as shown in the numerical results, can be very weakly underdamped, even in spite of the very strong system-coupling.}
    \label{fig:potential_well}
\end{figure*}

A possible interpretation of these results can be given following the analysis presented for the Rabi model in Ref.~\cite{Irish_Gea-Banacloche_2014}. There the authors consider a TLS coupled to a single quantum harmonic oscillator, with the total Hamiltonian of the TLS and the mode of the same form as our Hamiltonian \eref{eq:hamiltonian} (with a single mode in the bath); they moreover show that, whereas in the weak coupling limit the interaction with the system results in negligible displacement of the ground state of the harmonic oscillator, at strong coupling the lowest eigenstates of the overall TLS-mode Hamiltonian correspond to those of a particle in a (symmetric) double-well potential caused by strong environmental displacements.
In our setting, on the other side, we are dealing with a bath comprising an infinite number of harmonic modes. The results shown in Fig.~\ref{fig:spin_s=3} suggest however that, in the presence of a large enough system-bath coupling, a mechanism similar to the one discussed in Ref.~\cite{Irish_Gea-Banacloche_2014} is at play. This might be expected in the case of strongly super-Ohmic environments, as the spectral density becomes increasingly peaked around the cut-off frequency, leading to a quasi-monochromatic, underdamped bath correlation function with qualitative similarity to that of a single damped oscillator. To make this concrete, we can return to the chain Hamiltonian of the system (Eq. \ref{eq:chainHam}) and separate out the first bath mode ('the reaction coordinate' \cite{martinazzo2011communication,WoodsChin,iles2016energy,nazir2018reaction}) to obtain the following form 
\begin{align} \label{eq:reaction}
\hat{H}^C &= \hat{H}_S + g_0 \sigma_x (\hat{c}_0+\hat{c}_0^\dagger)+ \omega_0 \hat c_0^\dagger \hat c_0 \\
& +\sum_{k=1}^\infty \omega_j \hat c_k^\dagger \hat c_k + \sum_{k=1}^\infty g_k (\hat c_k^\dagger \hat{c}_{k-1} + \text{H.c.}). \nonumber
\end{align}
For sufficiently strong coupling (large $g_0$), we could consider the last two terms of the first line in Eq. \ref{eq:reaction} to be the reference Hamiltonian $H_0$, with $H_S$ and the terms of the second line acting as perturbations on the system and the collective reaction coordinate \cite{martinazzo2011communication,WoodsChin,iles2016energy,nazir2018reaction}, respectively. In this case, the exact (degenerate) groundstates of $H_0$ are given by $|\psi_\pm\rangle = |\sigma_{x=\pm 1}\rangle\otimes\exp(\mp g_0 \omega^{-1}(c_0 - c_0^\dagger))|0\rangle $, i.e. the mode is simply displaced in a direction that depends on the sign of the TLS projection along the $x$-axis. If we were to consider the energy of $H_0$ as a continuous function of a coherent state of the mode with displacement $q$, we would find a parabolic form with minima at $q=\mp g_0/\omega_0$ for spin states $|\sigma_{x=\pm 1}\rangle$. If we now reintroduce a small bare system Hamiltonian $H_S$ in this basis, we would expect energy gaps to emerge at the (now) avoided crossing of these parabolas, leading to a double 'potential' $V(q)$ for the TLS ,as sketched in Fig. \ref{fig:potential_well} (left).          

In the ultra-strong coupling regime considered here, the spin dynamics shown in Fig.~\ref{fig:spin_s=3} suggests that the presence of the TLS does indeed lead to the appearance of a double-well structure in $V(\hat{q})$, as depicted in Fig.~\ref{fig:potential_well}. At the minima of the two wells, the system eigenstates are no longer those of the bare TLS Hamiltonian ($\hat \sigma_z$) but rather those of the system-bath \emph{interaction} Hamiltonian ($\hat \sigma_x$). This explains the relaxation of $\langle \hat \sigma_z(t) \rangle$ to zero, rather than $\langle \hat \sigma_z(t) \rangle\rightarrow -1$, as might be expected at low temperatures. However, if the potential barrier between the two wells is low enough, a particle can tunnel from one well to the other so that the dressed ground state is in a superposition of the two wells.  In this configuration, the action of the free system Hamiltonian is to tunnel from one well to the other and the tunnelling rate will be  proportional to the (renormalized) TLS's frequency $\bar{\epsilon}_{\kappa=400}$. In the picture of Fig.\ref{fig:potential_well}, this renormalization can be seen to arise from the reduced overlap of the oscillator wave functions in each well: as the potential minima become mutually displaced, the probability to tunnel from one minimum to the other vanishes exponentially \cite{Chin_QPT_sub-ohmic_SBModel_ansatz,Leggett,blunden2017anatomy}.    

Taken together, the numerical results suggest a rich, multistage dynamics that are sketched in Fig. \ref{fig:potential_well}. Initially and ultra rapidly, the excited TLS decays onto the lower potential surface where the oscillator wave functions bifurcate and relax on timescales $\approx \omega_c ^{-1}$ to the minima of the potential, creating a `cat-like' superposition state. The `tunneling' provided by $H_S$ causes the lowest energy states of the system to be super positions of the localized dressed states, split by $\bar{\epsilon}$. The last stage of the dynamics is then the bath-induced relaxation (by emission into the bath) to the lowest of these delocalized states. This is particularly slow for strongly super-Ohmic baths, as the renormalization reduces the effective energy gap, and we would expect that the energy lost rate would be proportional to $\bar{\epsilon}^s$. This also explains the slower damping of the $s=3$ coherence, compared to $s=2$.  {Related results have very recently been discussed in Ref.~\cite{nacke23}, which presents a comprehensive numerical study of TLS dynamics in the presence of a zero-temperature super-Ohmic bath, albeit from a system-only perspective. From these data and analysis of the independent boson model, they identify an inverse timescale $\omega_c$ after which a significant slow down in TLS tunneling dynamics is observed which has a non-monotonic dependence on the bath exponent $s$. As super-Ohmic baths are among the most frequently encountered environments in solid state systems, understanding the phenomenology reported in Ref.~\cite{nacke23} in terms of explicit environmental observables would be a very interesting area for future T-TEDOPA studies.}

Finally, we note that the final, delocalized  `polaron' state is approximately given by $|gs\rangle \approx \ket{1}_x -\ket{0}_x$ \cite{blunden2017anatomy}. Due to the entanglement with the TLS, tracing over the TLS state leaves the environment in an equal mixture of coherent states with opposite displacements, although these states are overlapping (given that tunneling is still observed). The fact that the reduced density matrix contains oppositely displaced coherent states means that the average value of the displacement is zero, whereas the expectation value of quadratic operators is non zero. Thus, the appearance of multimode correlations in Fig. \ref{fig:correls=3} can be shown to be a consequence of the entangled (polaronic) ground state formed in the regime discussed in this section. Indeed using the approximation $|gs\rangle \approx \ket{1}_x -\ket{0}_x$, it is easy to show that $C(\omega, \omega')=\omega_c \frac{\sqrt{J(\omega)J(\omega')}}{\omega \omega'}$, which for $s=3$ gives $C(\omega, \omega')\propto \sqrt{\omega \omega'}/\omega_c$. This naturally explains the stronger signals at high frequency, as well as some of the other structures of the spectrum. The entanglement with the TLS is essential for this signature to appear, as it is equally straight forward to show that \emph{all} correlations vanish identically, if the environment were displaced in just one unique direction (simple product of coherent states).                   
{Also the seemingly discrete, grid-like pattern of correlations in Fig. \ref{fig:correls=3} can be motivated by the emergence of the polaronic dynamics sketched in Fig. \ref{fig:potential_well}. Since the reduced density matrix of the environment features oppositely displaced coherent states, modes across the entire frequency range are populated. As time progresses, these coherent states become increasingly displaced, involving a broader range of Fock states in the environment based on Poissonian statistics. In other words, we expect the ``grid" of TLS mediated correlations to become finer over time as the polarons are formed. Initially, the correlations are localized at the TLS frequency only, and then involve more discrete modes as time progresses. This is also visible in Fig. \ref{fig:s=3pops}, where the number of peaks in the occupations increases with progressing simulation time. Such a behavior is also present in the Independent Boson model (pure dephasing) \cite{breuer_petruccione}, which is analytically solvable and can be used as a benchmark test. }

\section{\label{sec:concl}Conclusion and Outlook}

In this work, we have gained new insights into how T-TEDOPA's pure wave function formalism addresses the dynamical nature of finite-temperature non-equilibrium steady states, and provided new ideas on how to exploit the extended environment of T-TEDOPA not only to gain computational advantage but also to compute physical observables. Our findings conclusively confirm the hypothesis of the thermal cycle in both the sub-Ohmic and super-Ohmic chain-mapped environments. This confirmation is based on the observation of the continuous extension of the occupied oscillators of the chain, which is shown to be non-pathological, and the corresponding unbounded growth of the emission/absorption peaks in the extended environment. Additionally, we provide a microscopic characterization of pure dephasing arising from the spectral density function in the sub-Ohmic environment. In the super-Ohmic environment, we reveal the thermal cycle as correlated emission into positive and negative frequency modes in the extended environment representation. To complete the description, we recover  the picture in the original environment, where we find that our results match the expected stationary occupations of the environmental physical modes. Furthermore, we report a novel case in which non-perturbative coupling between the TLS and its environment leads to the formation of polarons, and thus to a change in the ground state of the Spin Boson model. This change is evident in the system observables, where tunneling dynamics are observed, as well as in the correlations of the extended environment.

The approach introduced in this paper has a broad scope and can be exploited for the investigation of OQSs which have so far been mostly treated  perturbatively, e.g within the Markov approximation. The possibility to track the evolution of environmental degrees of freedom in the strong coupling and non pure-dephasing regime can be most relevant for the understanding of the fundamental mechanisms underpinning a larger class of open quantum system dynamics, with possible fundamental and technological impact. Solid-state implementations of qubits~\cite{metha16,gambetta17}, nanoscale quantum thermal machines \cite{esposito15,prior22}, sensing and metrology~\cite{smirne16}, energy-charge conversion and exciton transport in solid-state devices~\cite{ribeiro15,mitchison18}, or biological light harvesting complexes~\cite{huelga13,felipe22}, are typical instances in which deviations from a Lindbladian evolution can play a significant role. The methodology proposed in this work can provide, for example, a systematic tool for the definition of effective reduced environments, able to preserve the most relevant features of the environment while allowing for more efficient simulation of complex open quantum systems by classical means \cite{nazir18,lorenzoni23,tama18,mascherpa20}. Furthermore, the newfound ability to examine previously uncharted environmental modes might be instrumental in the domain of reservoir engineering \cite{metelmann, Mirrahimi_2014}, where the dynamics of an OQS are manipulated by tailoring its interaction with the environment. The demonstration of a strong system-bath coupling leading to polaronic coherent dynamics provides an example how a reservoir might be designed in order to induce longer coherence times in the TLS.

\begin{acknowledgments}

The authors thank Thibaut Lacroix, Davide Ferracin and Andrea Smirne for many interesting discussions. AR acknowledges support from the Erasmus+ Traineeship Program. AWC wishes to acknowledge support from ANR Project ACCEPT (Grant No. ANR-19-CE24-0028). DT acknowledges support from PSR UniMi 2022 initiative.

\end{acknowledgments}

\appendix

\section{\label{app:mps}Computing the time evolution with MPS methods}

In this Appendix we specify the tensor network time evolution techniques that we used to perform the simulations, and we provide detail on the choice of parameters.
Tensor network representations of many-body states \cite{Orus_2014, Schollwoeck_2011, Cirac_Perez-Garcia_Schuch_Verstraete_2021} naturally encode the information about the entanglement structure of a quantum state, and offer an efficient representation for weakly correlated states.
In particular, MPS enable to efficiently represent and to simulate the dynamics of one-dimensional problems: the finite temperature dynamics of the spin-boson model can be calculated using MPS thanks to the T-TEDOPA chain mapping. In fact, the chain-mapped spin-boson Hamiltonian can be straightforwardly written as a Matrix Product Operator (MPO), dictating the time evolution of the pure initial state of the system and of the environment $\ket{\psi(0)}_{SE}$, which has the following MPS representation in the diagrammatic notation:

\begin{equation}\label{eq:mps}
    \begin{tikzpicture}[scale=1.2,inner sep=1mm]
    
    \node (psi) at (-0.3, 0) {$\ket{\psi(0)}_{SE} = $};
    \node[tensor,draw=green!90,fill=green!35,thick, minimum width = 0.7cm, minimum height = 0.6cm] (1) at (1,0) {$S$};
    \node (1spin) at (1,-0.7) {$i_S$};
    \draw[-] (1)--(1spin) ;
    \foreach \i [evaluate={\j=int(\i-1)}] in {2,...,3} {
        \j = \i+1;
        \node[tensor,draw=red!40!blue!50,fill=red!15!blue!18, minimum width = 0.5cm, minimum height = 0.5cm] (\i) at (\i, 0) {$A_{\j}$};
        \node (\i spin) at (\i, -0.7) {$i_\j$};
        \draw[-] (\i) -- (\i spin);
    };
    \node (4) at (4, 0) {$...$};
         \node[tensor,draw=red!40!blue!50,fill=red!15!blue!18, minimum width = 0.5cm, minimum height = 0.5cm] (5) at (5, 0) {$A_N$};
         \node (5 spin) at (5, -0.7) {$i_N$};
        \draw[-] (5) -- (5 spin);
    \foreach \i in {1,...,4} {
        \pgfmathtruncatemacro{\iplusone}{\i + 1};
        \draw[-] (\i) -- (\iplusone);
    };
    
    \node[scale=1.] (chi) at (1.5,0.25) {$\chi_S$};
    \node[scale=1.] (chi) at (2.5,0.25) {$\chi_1$};
    \node[scale=1.] (chi) at (3.5,0.25) {$\chi_2$};
    \node[scale=1.] (chi) at (4.4,0.25) {$\chi_{N-1}$};
        
\end{tikzpicture} 
\end{equation}

The tensors $\{S, A_i\}$ are the sites of the MPS. Each site has a free leg that runs
over the values: $\{ i_k \}_{k=1}^d$. The dimension $d$ of the $k-$th leg is the number of values that the index $i_k$ can take. We set it to be $d = 10$, after having checked with preliminary trials that a larger dimension was not changing the results, but had a larger computational cost. In addition to that, the $k-$th site is connected to the tensors $A_{k-1}$ and $A_{k+1}$ through virtual legs, of dimensions  $\chi_{k-1}$ and $\chi_k$ respectively. They are known as bond dimensions, and they are correlated to the amount of entanglement in the MPS. The total number of complex coefficient that we need to store in order to represent an MPS is: $N \chi^2 d$. For the MPS representation to be efficient, bond dimensions (and therefore correlations) must be low. 

To compute the time evolution, we exploited the Dynamically evolving Time Dependent Variational Principle algorithm (DTDVP) \cite{dunnett2020dynamically, Dunnett_Chin_2021_evolving}. The choice of using DTDVP is advantageous in comparison to both the one-site (1TDVP) and two-site (2TDVP) version of TDVP \cite{haegeman_unifying, Lubich_Oseledets_Vandereycken_2015}. The main advantage of the one-site 1TDVP algorithm is that it preserves the unitarity of the MPS during the time evolution. Its main problem, conversely, is that the time evolution is constrained to happen on a manifold constituted by tensors of fixed bond dimension, a quantity closely related to the amount of entanglement in the MPS, and such a bond dimension has therefore to be fixed before the beginning of the time evolution. This strategy will necessarily be non optimal: the growth of the bond dimensions required to describe the quantum state should ideally mirror the entanglement growth induced by the time evolution. 2TDVP allows for such a dynamical growth of the bond dimensions, and therefore better describes the entanglement in the MPS. It suffers however of other drawbacks: first of all, a truncation error is introduced (by the means of an SVD decomposition), which entails a loss of unitarity of the time-evolved MPS. Moreover, 2TDVP has bad scaling properties with the size of the local dimensions of the MPS: this is a major issue when dealing with bosons, as it is the case in this work. The DTDVP algorithm \footnote{The numerical methods employed use the software packages available at: \url{https://github.com/angusdunnett/MPSDynamics}.} combines the best features of 1TDVP and 2TDVP: it preserves unitarity, it has the same scaling properties of 1TDVP, and it adapts the bond dimensions to the entanglement evolution at each site and at each time-step. DTDVP does not suffer from a truncation error, but introduces only a projection error. 

The DTDVP algorithm requires a maximal bond dimension to be specified, in our case $\chi_\text{max}=100$, and a precision, $p = 0.001$, defined using a convergence measure \cite{dunnett2020dynamically}.  

\section{The computational cost of T-TEDOPA}\label{app:cost}

\begin{figure}[!]
\includegraphics[width=0.35\textwidth]{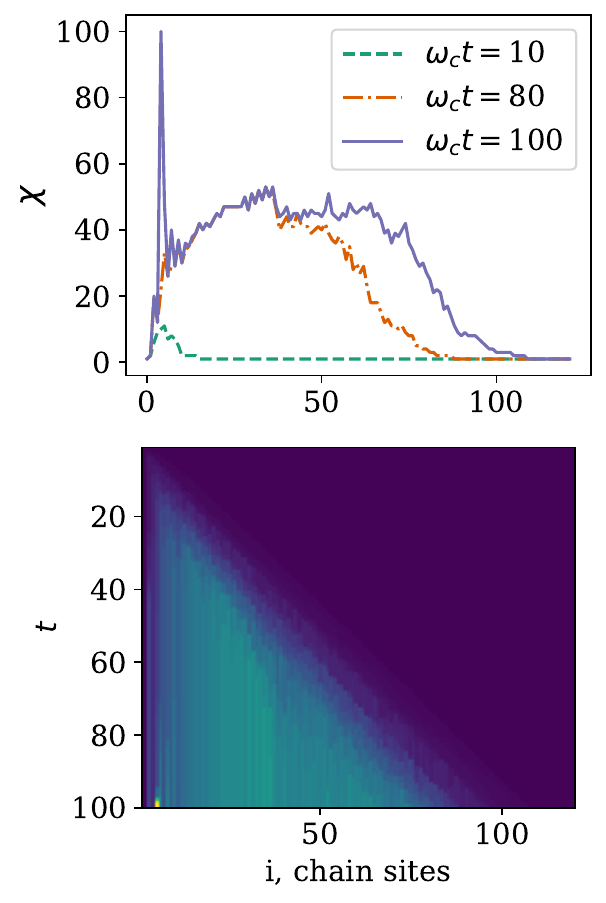}
\caption{\label{fig:bonds_evolution} Bond dimensions evolution for the super-Ohmic case. The inverse temperature is $\beta=2.0$, the initial state of the system is $\ket{1}$. In the top graph, the bond dimension $\chi$ is plotted against the chain site index, at different time-stamps. In the bottom graph, a heat-map of the bond dimension $\chi$ is shown in function of the chain site ($x-$axis) and of the simulation step ($y-$axis). }
\end{figure}

The T-TEDOPA approach gives the significant advantage of obtaining finite temperature dynamics with a pure state as the initial state of the time evolution. If instead the dynamics started from a thermal state:
\begin{equation}
    \hat \rho_\beta = \sum_{n_1 n_2 ...} e^{-\beta \sum_i n_i \omega_i } \ket{n_1 n_2 ...} \bra{n_1 n_2 ...}
\end{equation}
sampling the thermal probability distribution of the statistical ensemble of possible initial states would be intractably expensive. At the same time the computational cost for finite temperature simulations should increase. Tensor network methods work efficiently when the physical states represented by MPS are weakly correlated: the correlations between modes of opposite frequency introduced by the thermal cycle (between modes around zero frequency by the pure dephasing noise) will therefore cause a significant increase in the computational effort required to carry out the simulations. 

In fact, the correlations between the pairs of opposite frequency excitations will increase the entanglement in the chain modes, resulting in higher values for the bond dimensions $\chi_i$ of the system and the environment MPS (See Eq. \ref{eq:mps} in the Appendix \ref{app:mps}), and therefore the computational cost of the simulation. The higher the bond dimensions, the more the computational resources needed to store in memory and time evolve the MPS. The growth in computational cost mirrors the growth in entanglement that high bond dimension values represent. Thanks to time evolution algorithm used in our simulation (DTDVP) \cite{dunnett2020dynamically}, it is possible to follow the evolution of the bond dimensions at each chain site and at each simulation step, represented in Fig. \ref{fig:bonds_evolution} for the super-Ohmic, finite temperature case $\beta=2.0$. The maximum value reached by $\chi$ is for the first bond dimensions: a possible explanation could be that the correlations created in the thermal cycle, between positive and negative frequency modes, are mediated by the system. In accordance with expectations, higher temperatures correspond to higher entanglement in the MPS and hence a higher computational effort: in the low temperature case of $\beta=2000.0$, the maximum value reached by the bond dimensions is $\chi_\text{max}=7$.

\section{Remarks on the substitution of the coupling constant}\label{app:alpha}
In the super-Ohmic case, $s=2$,  the spectral density function has, in the considered interval $[-\omega_c,\omega_c]$, a lower value so that the coupling to the environmental modes is weaker than in the Ohmic case. This would translate in longer decay times in the super-Ohmic case; to make the decay rate independent of $s$, the value of the coupling strength $\alpha$ in Eq. \ref{eq:sdf_explicit} has been rescaled in the following way: 
\begin{equation}\label{alpha_s}
\alpha(s) = \frac{\alpha}{\epsilon^s}.    
\end{equation}
This rescaling makes the decay rate independent of $s$, thus allowing to have a constant decay rate for all of the $s$ values. In fact, the decay rate is a function of the spectral density function $J_s(\epsilon, \beta)$, proportional to $\alpha$, at the frequency value of the system $\omega=\epsilon$:
\begin{equation}
    J_s(\epsilon, \beta) = + \alpha \epsilon  \frac{\abs{\epsilon}^s}{\omega_c^{s-1}} \bigg( 1 + \coth\bigg( \frac{\beta \epsilon}{2}\bigg) \bigg) \theta(\abs{\epsilon}-\omega_c)
\end{equation}
The substitution $\alpha \to \alpha(s)$ makes the decay rate, and consequently the relaxation time $T_1$ defined in Sec. \ref{sec:results}, independent of the value of $s$, allowing for the observables to reach steady values in the same simulation time for any value of $s$.
The change of the coupling strength $\alpha(s)$ is reflected in the chain mapping as a change in the system-chain coupling  coefficient $g_0$ (see Eq.~\ref{eq:chainHam}) that has to be recalculated using the rescaled $\alpha_s$, whereas the other chain coefficients are left unchanged.

\section{\label{app:thermofield} The inverse thermofield transformation}

In the thermofield approach \cite{devegaban}, the first step is to introduce an auxiliary environment of non-interacting bosonic modes of negative frequencies:
\begin{equation}
    \hat H^\text{aux} = \hat H - \sum_{k=1}^N \omega_k \hat c_k^\dagger \hat c_k,
\end{equation}
where the Hamiltonian $\hat H$ is the bosonic Hamiltonian: 
\begin{equation}
    \hat H = \hat H_S + \sum_{k=1}^N \omega_k \hat b_k^\dagger \hat b_k + \frac{ \hat \sigma_x}{2} \otimes \sum_{k=1}^N g_k(\hat b_k^\dagger + \hat b_k).
\end{equation}
The two environments, of positive and negative frequencies, are assumed to be in a thermal environment at inverse temperature $\beta$; the second step is to apply a thermal Bogoliubov transformation to change the basis. The applied transformation produces two-modes squeezed states:
\begin{equation}\label{bogoliubov}
    \begin{split}
         &\hat a_{1k}=e^{-iG}  \hat b_k e^{iG}= \cosh(\theta_k) \hat b_k -\sinh(\theta_k)  \hat c_k^\dagger \\
         &\hat a_{2k}=e^{-iG} \hat c_k e^{iG}= \cosh(\theta_k)  \hat c_k -\sinh(\theta_k)  \hat b_k^\dagger,
    \end{split}
\end{equation}
where the exponent of the squeezing operator is $G = i \sum_k \theta_k(\hat b_k^\dagger \hat c_k^\dagger-\hat c_k \hat b_k)$, and $\theta_k$ is dependent on the temperature as in the following relations, where the average number of excitations in the $k$-th mode is $\langle \hat n_k \rangle = 1/(e^{\beta \omega_k}-1)$:
\begin{equation}\label{eq:hyperbolic}
    \begin{split}
         &\cosh(\theta_k) = \sqrt{1+\langle \hat n_k \rangle } = \sqrt{\frac{1}{1-e^{-\beta \omega_k}}} \\
         &\sinh(\theta_k) =\quad\sqrt{\langle \hat n_k \rangle }\quad= \sqrt{\frac{1}{e^{+\beta \omega_k}-1}}.
    \end{split}
\end{equation}
The Bogoliubov transformation defines a new squeezed vacuum state, which we write in terms of the vacuum state $\ket{\Omega_0}$ of the bosonic modes (original and auxiliary) of the operators $\hat b_k$, $\hat c_k$:
\begin{equation}
    \ket{\Omega} = e^{iG} \ket{\Omega_0}, \quad \hat a_{1k} \ket{\Omega} = 0,  \hat a_{2k} \ket{\Omega} = 0.
\end{equation}
From the vacuum state, we can obtain the thermal state of the original environment:
\begin{equation}
    \hat \rho_E = \Tr_{\text{aux}}\{\ket{\Omega}\bra{\Omega} \},
\end{equation}
and it can be now used as pure an initial state for both of the environments.

We now explain how to invert the thermofield transformation. The matrix $M$ defining the transformation of Eq. \ref{bogoliubov} is invertible: it is therefore possible to obtain the modes of the original ($\hat b_{k}$) and auxiliary ($\hat c_{k}$) environments as linear combination of the Bogoliubov-transformed modes $\hat a_{1k}$ and $\hat a_{2k}$. We explicitly write the transformation:
\begin{equation}
    \begin{split}
        \begin{bmatrix}
            \hat a_{1k} \\ \hat a_{2k} \\ \hat a_{1k}^\dagger \\ \hat a_{2k}^\dagger
        \end{bmatrix} = 
        \begin{bmatrix}
            \cosh{\theta_k} & 0 & 0 & -\sinh{\theta_k} \\ 
            0 & \cosh{\theta_k} & -\sinh{\theta_k} & 0 \\ 
            0 & -\sinh{\theta_k} & \cosh{\theta_k} & 0 \\ 
            -\sinh{\theta_k} & 0 & 0 & \cosh{\theta_k}
        \end{bmatrix} 
        \begin{bmatrix}
            \hat b_{k} \\ \hat c_{k} \\ \hat b_{k}^\dagger \\ \hat c_{k}^\dagger
        \end{bmatrix}, 
    \end{split}
\end{equation}
and its inverse:
\begin{equation}
    \begin{split}
        \begin{bmatrix}
            \hat b_{k} \\ \hat c_{k} \\ \hat b_{k}^\dagger \\ \hat c_{k}^\dagger
        \end{bmatrix} = 
        \begin{bmatrix}
            \cosh{\theta_k} & 0 & 0 & \sinh{\theta_k} \\ 
            0 & \cosh{\theta_k} & \sinh{\theta_k} & 0 \\ 
            0 & \sinh{\theta_k} & \cosh{\theta_k} & 0 \\ 
            \sinh{\theta_k} & 0 & 0 & \cosh{\theta_k}
        \end{bmatrix} 
        \begin{bmatrix}
            \hat a_{1k} \\ \hat a_{2k} \\ \hat a_{1k}^\dagger \\ \hat a_{2k}^\dagger
        \end{bmatrix}.
    \end{split}
\end{equation}
It is now possible to obtain the mean value of the number operator for the physical modes as a function of mean values that we already have evaluated:

\begin{eqnarray}\label{eq:physical_occupations}
    \langle \hat b_k^\dagger \hat b_k \rangle =& \cosh{\theta_k}\sinh{\theta_k} \Big(\langle\hat a_{2k}\hat a_{1k}\rangle + \langle \hat a_{1k}^\dagger\hat a_{2k}^\dagger\rangle \Big)+ \\ &+ \sinh^2{\theta_k} \Big(1+ \langle \hat a_{2k}^\dagger \hat a_{2k} \rangle \Big) + \cosh^2{\theta_k} \langle \hat a_{1k}^\dagger \hat a_{1k}, \rangle \nonumber
\end{eqnarray}
where we exploited the bosonic commutation relations to obtain: $\hat a_{1k} \hat a_{1k}^\dagger = 1 + \hat a_{1k}^\dagger \hat a_{1k}$. In the thermofield case, a negative frequency $\omega_{2k}$ is associated to each positive frequency $\omega_{1k}$. The sampling is therefore symmetric around zero. This marks a difference with T-TEDOPA, where the sampling of frequencies is obtained through the thermalized measure $d\mu(\beta) = \sqrt{J(\omega, \beta)}d\omega$, and is not symmetric. To recover the results for the physical bath of frequencies starting from the results of our simulations, conducted using the T-TEDOPA chain mapping, we need to do an extrapolation for all of the mean values appearing in Eq. \ref{eq:physical_occupations}, in order to have their values for each $\omega$ at $-\omega$ as well. Given this need, the enlarged grid of frequencies (Appendix \ref{app:grid}) is particularly useful. To sum up: from our simulations we obtained the occupations of the chain modes $\hat c_k$; we inverted the unitary transformation that defines the chain mapping (as in Eq. \ref{eq:chain_modes}) to obtain the modes in the extended bath $\hat n_i^b$; we inverted the thermofield transformation $M$ to obtain the physical modes occupations in the bath of positive frequencies.

\section{\label{app:grid}Technical observation on the frequency sampling}

As a final technical observation, we note that by exploiting the wave-like propagation of the chain modes during the dynamics, and the analytical properties of the chain mapping, it is possible to calculate both the occupations and the correlators of the modes in the extended bath for a finer grid of frequency values, instead of only using the $N=120$ chain modes. The excitations propagate on the chain as waves: if the wave has not reached the $i$ site, the site $i+1$ will have zero as the occupation value $\langle \hat c^\dagger \hat c \rangle= 0$. Therefore, we calculate the chain coefficients for a chain of $M=700$ sites, and we artificially enlarge, with $0$s as entries, the matrices with the expectation values of the chain modes creation and annihilation operators on the time evolved state of the system and the environment. For example, blocks of zeros are added to the $N\times N$ matrix $N_{ij}=[\langle \hat c_i^\dagger \hat c_i \rangle ]$, so that it becomes an $M \times M$ matrix. It is reasonable to do so: it is as if the dynamics was happening on a longer chain, where the modes not yet reached by the excitation's wave stay unoccupied. On the corresponding sites, $N<i<M$, the chain modes creation and annihilation operators will have zero as an expectation value. A longer chain, as explained in Sec. \ref{sec:methods}, will result in a better sampling in the extended bath domain. Once we obtain the frequency sampling $\{\omega_1, \omega_2, ... \}$, the spectral density $J_\beta(\omega_i)$ must be weighted by the corresponding frequency interval $[\omega_{i+1}, \omega_i]$.


\begin{thebibliography}{63}%
\makeatletter
\providecommand \@ifxundefined [1]{%
 \@ifx{#1\undefined}
}%
\providecommand \@ifnum [1]{%
 \ifnum #1\expandafter \@firstoftwo
 \else \expandafter \@secondoftwo
 \fi
}%
\providecommand \@ifx [1]{%
 \ifx #1\expandafter \@firstoftwo
 \else \expandafter \@secondoftwo
 \fi
}%
\providecommand \natexlab [1]{#1}%
\providecommand \enquote  [1]{``#1''}%
\providecommand \bibnamefont  [1]{#1}%
\providecommand \bibfnamefont [1]{#1}%
\providecommand \citenamefont [1]{#1}%
\providecommand \href@noop [0]{\@secondoftwo}%
\providecommand \href [0]{\begingroup \@sanitize@url \@href}%
\providecommand \@href[1]{\@@startlink{#1}\@@href}%
\providecommand \@@href[1]{\endgroup#1\@@endlink}%
\providecommand \@sanitize@url [0]{\catcode `\\12\catcode `\$12\catcode
  `\&12\catcode `\#12\catcode `\^12\catcode `\_12\catcode `\%12\relax}%
\providecommand \@@startlink[1]{}%
\providecommand \@@endlink[0]{}%
\providecommand \url  [0]{\begingroup\@sanitize@url \@url }%
\providecommand \@url [1]{\endgroup\@href {#1}{\urlprefix }}%
\providecommand \urlprefix  [0]{URL }%
\providecommand \Eprint [0]{\href }%
\providecommand \doibase [0]{https://doi.org/}%
\providecommand \selectlanguage [0]{\@gobble}%
\providecommand \bibinfo  [0]{\@secondoftwo}%
\providecommand \bibfield  [0]{\@secondoftwo}%
\providecommand \translation [1]{[#1]}%
\providecommand \BibitemOpen [0]{}%
\providecommand \bibitemStop [0]{}%
\providecommand \bibitemNoStop [0]{.\EOS\space}%
\providecommand \EOS [0]{\spacefactor3000\relax}%
\providecommand \BibitemShut  [1]{\csname bibitem#1\endcsname}%
\let\auto@bib@innerbib\@empty
\bibitem [{\citenamefont {Breuer}\ and\ \citenamefont
  {Petruccione}(2002)}]{breuer_petruccione}%
  \BibitemOpen
  \bibfield  {author} {\bibinfo {author} {\bibfnamefont {H.~P.}\ \bibnamefont
  {Breuer}}\ and\ \bibinfo {author} {\bibfnamefont {F.}~\bibnamefont
  {Petruccione}},\ }\href@noop {} {\emph {\bibinfo {title} {The theory of open
  quantum systems}}}\ (\bibinfo  {publisher} {Oxford University Press},\
  \bibinfo {address} {Great Clarendon Street},\ \bibinfo {year}
  {2002})\BibitemShut {NoStop}%
\bibitem [{\citenamefont {Rivas}\ and\ \citenamefont
  {Huelga}(2012)}]{rivas2012open}%
  \BibitemOpen
  \bibfield  {author} {\bibinfo {author} {\bibfnamefont {A.}~\bibnamefont
  {Rivas}}\ and\ \bibinfo {author} {\bibfnamefont {S.~F.}\ \bibnamefont
  {Huelga}},\ }\href@noop {} {\emph {\bibinfo {title} {Open quantum
  systems}}},\ Vol.~\bibinfo {volume} {10}\ (\bibinfo  {publisher} {Springer},\
  \bibinfo {year} {2012})\BibitemShut {NoStop}%
\bibitem [{\citenamefont {Leggett}\ \emph {et~al.}(1987)\citenamefont
  {Leggett}, \citenamefont {Chakravarty}, \citenamefont {Dorsey}, \citenamefont
  {Fisher}, \citenamefont {Garg},\ and\ \citenamefont {Zwerger}}]{Leggett}%
  \BibitemOpen
  \bibfield  {author} {\bibinfo {author} {\bibfnamefont {A.~J.}\ \bibnamefont
  {Leggett}}, \bibinfo {author} {\bibfnamefont {S.}~\bibnamefont
  {Chakravarty}}, \bibinfo {author} {\bibfnamefont {A.~T.}\ \bibnamefont
  {Dorsey}}, \bibinfo {author} {\bibfnamefont {M.~P.~A.}\ \bibnamefont
  {Fisher}}, \bibinfo {author} {\bibfnamefont {A.}~\bibnamefont {Garg}},\ and\
  \bibinfo {author} {\bibfnamefont {W.}~\bibnamefont {Zwerger}},\ }\bibfield
  {title} {\bibinfo {title} {Dynamics of the dissipative two-state system},\
  }\href {https://doi.org/10.1103/RevModPhys.59.1} {\bibfield  {journal}
  {\bibinfo  {journal} {Rev. Mod. Phys.}\ }\textbf {\bibinfo {volume} {59}},\
  \bibinfo {pages} {1} (\bibinfo {year} {1987})}\BibitemShut {NoStop}%
\bibitem [{\citenamefont {Rotter}\ and\ \citenamefont
  {Bird}(2015)}]{rotter2015review}%
  \BibitemOpen
  \bibfield  {author} {\bibinfo {author} {\bibfnamefont {I.}~\bibnamefont
  {Rotter}}\ and\ \bibinfo {author} {\bibfnamefont {J.}~\bibnamefont {Bird}},\
  }\bibfield  {title} {\bibinfo {title} {A review of progress in the physics of
  open quantum systems: theory and experiment},\ }\href@noop {} {\bibfield
  {journal} {\bibinfo  {journal} {Reports on Progress in Physics}\ }\textbf
  {\bibinfo {volume} {78}},\ \bibinfo {pages} {114001} (\bibinfo {year}
  {2015})}\BibitemShut {NoStop}%
\bibitem [{\citenamefont {Koch}\ \emph {et~al.}(2022)\citenamefont {Koch},
  \citenamefont {Boscain}, \citenamefont {Calarco}, \citenamefont {Dirr},
  \citenamefont {Filipp}, \citenamefont {Glaser}, \citenamefont {Kosloff},
  \citenamefont {Montangero}, \citenamefont {Schulte-Herbr{\"u}ggen},
  \citenamefont {Sugny} \emph {et~al.}}]{koch2022quantum}%
  \BibitemOpen
  \bibfield  {author} {\bibinfo {author} {\bibfnamefont {C.~P.}\ \bibnamefont
  {Koch}}, \bibinfo {author} {\bibfnamefont {U.}~\bibnamefont {Boscain}},
  \bibinfo {author} {\bibfnamefont {T.}~\bibnamefont {Calarco}}, \bibinfo
  {author} {\bibfnamefont {G.}~\bibnamefont {Dirr}}, \bibinfo {author}
  {\bibfnamefont {S.}~\bibnamefont {Filipp}}, \bibinfo {author} {\bibfnamefont
  {S.~J.}\ \bibnamefont {Glaser}}, \bibinfo {author} {\bibfnamefont
  {R.}~\bibnamefont {Kosloff}}, \bibinfo {author} {\bibfnamefont
  {S.}~\bibnamefont {Montangero}}, \bibinfo {author} {\bibfnamefont
  {T.}~\bibnamefont {Schulte-Herbr{\"u}ggen}}, \bibinfo {author} {\bibfnamefont
  {D.}~\bibnamefont {Sugny}}, \emph {et~al.},\ }\bibfield  {title} {\bibinfo
  {title} {Quantum optimal control in quantum technologies. strategic report on
  current status, visions and goals for research in europe},\ }\href@noop {}
  {\bibfield  {journal} {\bibinfo  {journal} {EPJ Quantum Technology}\ }\textbf
  {\bibinfo {volume} {9}},\ \bibinfo {pages} {19} (\bibinfo {year}
  {2022})}\BibitemShut {NoStop}%
\bibitem [{\citenamefont {Auffeves}(2022)}]{auffeves2022quantum}%
  \BibitemOpen
  \bibfield  {author} {\bibinfo {author} {\bibfnamefont {A.}~\bibnamefont
  {Auffeves}},\ }\bibfield  {title} {\bibinfo {title} {Quantum technologies
  need a quantum energy initiative},\ }\href@noop {} {\bibfield  {journal}
  {\bibinfo  {journal} {PRX Quantum}\ }\textbf {\bibinfo {volume} {3}},\
  \bibinfo {pages} {020101} (\bibinfo {year} {2022})}\BibitemShut {NoStop}%
\bibitem [{\citenamefont {Farina}\ \emph {et~al.}(2019)\citenamefont {Farina},
  \citenamefont {Andolina}, \citenamefont {Mari}, \citenamefont {Polini},\ and\
  \citenamefont {Giovannetti}}]{farina2019charger}%
  \BibitemOpen
  \bibfield  {author} {\bibinfo {author} {\bibfnamefont {D.}~\bibnamefont
  {Farina}}, \bibinfo {author} {\bibfnamefont {G.~M.}\ \bibnamefont
  {Andolina}}, \bibinfo {author} {\bibfnamefont {A.}~\bibnamefont {Mari}},
  \bibinfo {author} {\bibfnamefont {M.}~\bibnamefont {Polini}},\ and\ \bibinfo
  {author} {\bibfnamefont {V.}~\bibnamefont {Giovannetti}},\ }\bibfield
  {title} {\bibinfo {title} {Charger-mediated energy transfer for quantum
  batteries: An open-system approach},\ }\href@noop {} {\bibfield  {journal}
  {\bibinfo  {journal} {Physical Review B}\ }\textbf {\bibinfo {volume} {99}},\
  \bibinfo {pages} {035421} (\bibinfo {year} {2019})}\BibitemShut {NoStop}%
\bibitem [{\citenamefont {Wertnik}\ \emph {et~al.}(2018)\citenamefont
  {Wertnik}, \citenamefont {Chin}, \citenamefont {Nori},\ and\ \citenamefont
  {Lambert}}]{wertnik2018optimizing}%
  \BibitemOpen
  \bibfield  {author} {\bibinfo {author} {\bibfnamefont {M.}~\bibnamefont
  {Wertnik}}, \bibinfo {author} {\bibfnamefont {A.}~\bibnamefont {Chin}},
  \bibinfo {author} {\bibfnamefont {F.}~\bibnamefont {Nori}},\ and\ \bibinfo
  {author} {\bibfnamefont {N.}~\bibnamefont {Lambert}},\ }\bibfield  {title}
  {\bibinfo {title} {Optimizing co-operative multi-environment dynamics in a
  dark-state-enhanced photosynthetic heat engine},\ }\href@noop {} {\bibfield
  {journal} {\bibinfo  {journal} {The Journal of chemical physics}\ }\textbf
  {\bibinfo {volume} {149}},\ \bibinfo {pages} {084112} (\bibinfo {year}
  {2018})}\BibitemShut {NoStop}%
\bibitem [{\citenamefont {Alvertis}\ \emph {et~al.}(2020)\citenamefont
  {Alvertis}, \citenamefont {Pandya}, \citenamefont {Muscarella}, \citenamefont
  {Sawhney}, \citenamefont {Nguyen}, \citenamefont {Ehrler}, \citenamefont
  {Rao}, \citenamefont {Friend}, \citenamefont {Chin},\ and\ \citenamefont
  {Monserrat}}]{alvertis2020impact}%
  \BibitemOpen
  \bibfield  {author} {\bibinfo {author} {\bibfnamefont {A.~M.}\ \bibnamefont
  {Alvertis}}, \bibinfo {author} {\bibfnamefont {R.}~\bibnamefont {Pandya}},
  \bibinfo {author} {\bibfnamefont {L.~A.}\ \bibnamefont {Muscarella}},
  \bibinfo {author} {\bibfnamefont {N.}~\bibnamefont {Sawhney}}, \bibinfo
  {author} {\bibfnamefont {M.}~\bibnamefont {Nguyen}}, \bibinfo {author}
  {\bibfnamefont {B.}~\bibnamefont {Ehrler}}, \bibinfo {author} {\bibfnamefont
  {A.}~\bibnamefont {Rao}}, \bibinfo {author} {\bibfnamefont {R.~H.}\
  \bibnamefont {Friend}}, \bibinfo {author} {\bibfnamefont {A.~W.}\
  \bibnamefont {Chin}},\ and\ \bibinfo {author} {\bibfnamefont
  {B.}~\bibnamefont {Monserrat}},\ }\bibfield  {title} {\bibinfo {title}
  {Impact of exciton delocalization on exciton-vibration interactions in
  organic semiconductors},\ }\href@noop {} {\bibfield  {journal} {\bibinfo
  {journal} {Physical Review B}\ }\textbf {\bibinfo {volume} {102}},\ \bibinfo
  {pages} {081122} (\bibinfo {year} {2020})}\BibitemShut {NoStop}%
\bibitem [{\citenamefont {Dunnett}\ \emph {et~al.}(2021)\citenamefont
  {Dunnett}, \citenamefont {Gowland}, \citenamefont {Isborn}, \citenamefont
  {Chin},\ and\ \citenamefont {Zuehlsdorff}}]{dunnett2021influence}%
  \BibitemOpen
  \bibfield  {author} {\bibinfo {author} {\bibfnamefont {A.~J.}\ \bibnamefont
  {Dunnett}}, \bibinfo {author} {\bibfnamefont {D.}~\bibnamefont {Gowland}},
  \bibinfo {author} {\bibfnamefont {C.~M.}\ \bibnamefont {Isborn}}, \bibinfo
  {author} {\bibfnamefont {A.~W.}\ \bibnamefont {Chin}},\ and\ \bibinfo
  {author} {\bibfnamefont {T.~J.}\ \bibnamefont {Zuehlsdorff}},\ }\bibfield
  {title} {\bibinfo {title} {Influence of non-adiabatic effects on linear
  absorption spectra in the condensed phase: Methylene blue},\ }\href@noop {}
  {\bibfield  {journal} {\bibinfo  {journal} {The journal of chemical physics}\
  }\textbf {\bibinfo {volume} {155}},\ \bibinfo {pages} {144112} (\bibinfo
  {year} {2021})}\BibitemShut {NoStop}%
\bibitem [{\citenamefont {Schollwoeck}(2011)}]{Schollwoeck_2011}%
  \BibitemOpen
  \bibfield  {author} {\bibinfo {author} {\bibfnamefont {U.}~\bibnamefont
  {Schollwoeck}},\ }\bibfield  {title} {\bibinfo {title} {The density-matrix
  renormalization group in the age of matrix product states},\ }\href
  {https://doi.org/10.1016/j.aop.2010.09.012} {\bibfield  {journal} {\bibinfo
  {journal} {Annals of Physics}\ }\textbf {\bibinfo {volume} {326}},\ \bibinfo
  {pages} {96–192} (\bibinfo {year} {2011})},\ \bibinfo {note} {arXiv:
  1008.3477}\BibitemShut {NoStop}%
\bibitem [{\citenamefont {Orus}(2014)}]{Orus_2014}%
  \BibitemOpen
  \bibfield  {author} {\bibinfo {author} {\bibfnamefont {R.}~\bibnamefont
  {Orus}},\ }\bibfield  {title} {\bibinfo {title} {A practical introduction to
  tensor networks: Matrix product states and projected entangled pair states},\
  }\bibfield  {journal} {\bibinfo  {journal} {arXiv:1306.2164 [cond-mat,
  physics:hep-lat, physics:hep-th, physics:quant-ph]}\ }\href
  {https://doi.org/10.1016/j.aop.2014.06.013} {10.1016/j.aop.2014.06.013}
  (\bibinfo {year} {2014}),\ \bibinfo {note} {arXiv: 1306.2164}\BibitemShut
  {NoStop}%
\bibitem [{\citenamefont {Vidal}(2003)}]{vidal03}%
  \BibitemOpen
  \bibfield  {author} {\bibinfo {author} {\bibfnamefont {G.}~\bibnamefont
  {Vidal}},\ }\bibfield  {title} {\bibinfo {title} {Efficient classical
  simulation of slightly entangled quantum computations},\ }\href@noop {}
  {\bibfield  {journal} {\bibinfo  {journal} {Phys. Rev. Lett.}\ }\textbf
  {\bibinfo {volume} {91}},\ \bibinfo {pages} {147902} (\bibinfo {year}
  {2003})}\BibitemShut {NoStop}%
\bibitem [{\citenamefont {Vidal}(2004)}]{vidal04}%
  \BibitemOpen
  \bibfield  {author} {\bibinfo {author} {\bibfnamefont {G.}~\bibnamefont
  {Vidal}},\ }\bibfield  {title} {\bibinfo {title} {Efficient simulation of
  one-dimensional quantum many-body systems},\ }\href@noop {} {\bibfield
  {journal} {\bibinfo  {journal} {Phys. Rev. Lett.}\ }\textbf {\bibinfo
  {volume} {93}},\ \bibinfo {pages} {040502} (\bibinfo {year}
  {2004})}\BibitemShut {NoStop}%
\bibitem [{\citenamefont {Daley}\ \emph {et~al.}(2004)\citenamefont {Daley},
  \citenamefont {Kollath}, \citenamefont {Schollwöck},\ and\ \citenamefont
  {Vidal}}]{daley04}%
  \BibitemOpen
  \bibfield  {author} {\bibinfo {author} {\bibfnamefont {A.~J.}\ \bibnamefont
  {Daley}}, \bibinfo {author} {\bibfnamefont {C.}~\bibnamefont {Kollath}},
  \bibinfo {author} {\bibfnamefont {U.}~\bibnamefont {Schollwöck}},\ and\
  \bibinfo {author} {\bibfnamefont {G.}~\bibnamefont {Vidal}},\ }\bibfield
  {title} {\bibinfo {title} {Time-dependent density-matrix
  renormalization-group using adaptive effective hilbert spaces},\ }\href
  {https://doi.org/10.1088/1742-5468/2004/04/P04005} {\bibfield  {journal}
  {\bibinfo  {journal} {Journal of Statistical Mechanics: Theory and
  Experiment}\ }\textbf {\bibinfo {volume} {2004}},\ \bibinfo {pages} {P04005}
  (\bibinfo {year} {2004})}\BibitemShut {NoStop}%
\bibitem [{\citenamefont {Zwolak}\ and\ \citenamefont
  {Vidal}(2004)}]{zwolak04}%
  \BibitemOpen
  \bibfield  {author} {\bibinfo {author} {\bibfnamefont {M.}~\bibnamefont
  {Zwolak}}\ and\ \bibinfo {author} {\bibfnamefont {G.}~\bibnamefont {Vidal}},\
  }\bibfield  {title} {\bibinfo {title} {Mixed-state dynamics in
  one-dimensional quantum lattice systems: A time-dependent superoperator
  renormalization algorithm},\ }\href@noop {} {\bibfield  {journal} {\bibinfo
  {journal} {Phys. Rev. Lett.}\ }\textbf {\bibinfo {volume} {93}},\ \bibinfo
  {pages} {207205} (\bibinfo {year} {2004})}\BibitemShut {NoStop}%
\bibitem [{\citenamefont {Haegeman}\ \emph {et~al.}(2016)\citenamefont
  {Haegeman}, \citenamefont {Lubich}, \citenamefont {Oseledets}, \citenamefont
  {Vandereycken},\ and\ \citenamefont {Verstraete}}]{haegeman_unifying}%
  \BibitemOpen
  \bibfield  {author} {\bibinfo {author} {\bibfnamefont {J.}~\bibnamefont
  {Haegeman}}, \bibinfo {author} {\bibfnamefont {C.}~\bibnamefont {Lubich}},
  \bibinfo {author} {\bibfnamefont {I.}~\bibnamefont {Oseledets}}, \bibinfo
  {author} {\bibfnamefont {B.}~\bibnamefont {Vandereycken}},\ and\ \bibinfo
  {author} {\bibfnamefont {F.}~\bibnamefont {Verstraete}},\ }\bibfield  {title}
  {\bibinfo {title} {Unifying time evolution and optimization with matrix
  product states},\ }\href {https://doi.org/10.1103/PhysRevB.94.165116}
  {\bibfield  {journal} {\bibinfo  {journal} {Phys. Rev. B}\ }\textbf {\bibinfo
  {volume} {94}},\ \bibinfo {pages} {165116} (\bibinfo {year}
  {2016})}\BibitemShut {NoStop}%
\bibitem [{\citenamefont {Lubich}\ \emph {et~al.}(2015)\citenamefont {Lubich},
  \citenamefont {Oseledets},\ and\ \citenamefont
  {Vandereycken}}]{Lubich_Oseledets_Vandereycken_2015}%
  \BibitemOpen
  \bibfield  {author} {\bibinfo {author} {\bibfnamefont {C.}~\bibnamefont
  {Lubich}}, \bibinfo {author} {\bibfnamefont {I.~V.}\ \bibnamefont
  {Oseledets}},\ and\ \bibinfo {author} {\bibfnamefont {B.}~\bibnamefont
  {Vandereycken}},\ }\bibfield  {title} {\bibinfo {title} {Time integration of
  tensor trains},\ }\href {https://doi.org/10.1137/140976546} {\bibfield
  {journal} {\bibinfo  {journal} {SIAM Journal on Numerical Analysis}\ }\textbf
  {\bibinfo {volume} {53}},\ \bibinfo {pages} {917–941} (\bibinfo {year}
  {2015})}\BibitemShut {NoStop}%
\bibitem [{\citenamefont {Prior}\ \emph {et~al.}(2010)\citenamefont {Prior},
  \citenamefont {Chin}, \citenamefont {Huelga},\ and\ \citenamefont
  {Plenio}}]{prior2010efficient}%
  \BibitemOpen
  \bibfield  {author} {\bibinfo {author} {\bibfnamefont {J.}~\bibnamefont
  {Prior}}, \bibinfo {author} {\bibfnamefont {A.~W.}\ \bibnamefont {Chin}},
  \bibinfo {author} {\bibfnamefont {S.~F.}\ \bibnamefont {Huelga}},\ and\
  \bibinfo {author} {\bibfnamefont {M.~B.}\ \bibnamefont {Plenio}},\ }\bibfield
   {title} {\bibinfo {title} {Efficient simulation of strong system-environment
  interactions},\ }\href@noop {} {\bibfield  {journal} {\bibinfo  {journal}
  {Physical review letters}\ }\textbf {\bibinfo {volume} {105}},\ \bibinfo
  {pages} {050404} (\bibinfo {year} {2010})}\BibitemShut {NoStop}%
\bibitem [{\citenamefont {Oviedo-Casado}\ \emph {et~al.}(2016)\citenamefont
  {Oviedo-Casado}, \citenamefont {Prior}, \citenamefont {Chin}, \citenamefont
  {Rosenbach}, \citenamefont {Huelga},\ and\ \citenamefont
  {Plenio}}]{oviedo2016phase}%
  \BibitemOpen
  \bibfield  {author} {\bibinfo {author} {\bibfnamefont {S.}~\bibnamefont
  {Oviedo-Casado}}, \bibinfo {author} {\bibfnamefont {J.}~\bibnamefont
  {Prior}}, \bibinfo {author} {\bibfnamefont {A.}~\bibnamefont {Chin}},
  \bibinfo {author} {\bibfnamefont {R.}~\bibnamefont {Rosenbach}}, \bibinfo
  {author} {\bibfnamefont {S.}~\bibnamefont {Huelga}},\ and\ \bibinfo {author}
  {\bibfnamefont {M.}~\bibnamefont {Plenio}},\ }\bibfield  {title} {\bibinfo
  {title} {Phase-dependent exciton transport and energy harvesting from thermal
  environments},\ }\href@noop {} {\bibfield  {journal} {\bibinfo  {journal}
  {Physical Review A}\ }\textbf {\bibinfo {volume} {93}},\ \bibinfo {pages}
  {020102} (\bibinfo {year} {2016})}\BibitemShut {NoStop}%
\bibitem [{\citenamefont {Del~Pino}\ \emph {et~al.}(2018)\citenamefont
  {Del~Pino}, \citenamefont {Schr{\"o}der}, \citenamefont {Chin}, \citenamefont
  {Feist},\ and\ \citenamefont {Garcia-Vidal}}]{del2018tensor}%
  \BibitemOpen
  \bibfield  {author} {\bibinfo {author} {\bibfnamefont {J.}~\bibnamefont
  {Del~Pino}}, \bibinfo {author} {\bibfnamefont {F.~A.}\ \bibnamefont
  {Schr{\"o}der}}, \bibinfo {author} {\bibfnamefont {A.~W.}\ \bibnamefont
  {Chin}}, \bibinfo {author} {\bibfnamefont {J.}~\bibnamefont {Feist}},\ and\
  \bibinfo {author} {\bibfnamefont {F.~J.}\ \bibnamefont {Garcia-Vidal}},\
  }\bibfield  {title} {\bibinfo {title} {Tensor network simulation of
  polaron-polaritons in organic microcavities},\ }\href@noop {} {\bibfield
  {journal} {\bibinfo  {journal} {Physical Review B}\ }\textbf {\bibinfo
  {volume} {98}},\ \bibinfo {pages} {165416} (\bibinfo {year}
  {2018})}\BibitemShut {NoStop}%
\bibitem [{\citenamefont {Tamascelli}\ \emph {et~al.}(2019)\citenamefont
  {Tamascelli}, \citenamefont {Smirne}, \citenamefont {Lim}, \citenamefont
  {Huelga},\ and\ \citenamefont {Plenio}}]{Tamascelli_ttedopa}%
  \BibitemOpen
  \bibfield  {author} {\bibinfo {author} {\bibfnamefont {D.}~\bibnamefont
  {Tamascelli}}, \bibinfo {author} {\bibfnamefont {A.}~\bibnamefont {Smirne}},
  \bibinfo {author} {\bibfnamefont {J.}~\bibnamefont {Lim}}, \bibinfo {author}
  {\bibfnamefont {S.~F.}\ \bibnamefont {Huelga}},\ and\ \bibinfo {author}
  {\bibfnamefont {M.~B.}\ \bibnamefont {Plenio}},\ }\bibfield  {title}
  {\bibinfo {title} {Efficient simulation of finite-temperature open quantum
  systems},\ }\href {https://doi.org/10.1103/PhysRevLett.123.090402} {\bibfield
   {journal} {\bibinfo  {journal} {Phys. Rev. Lett.}\ }\textbf {\bibinfo
  {volume} {123}},\ \bibinfo {pages} {090402} (\bibinfo {year}
  {2019})}\BibitemShut {NoStop}%
\bibitem [{\citenamefont {Tamascelli}(2020)}]{TAMA2020}%
  \BibitemOpen
  \bibfield  {author} {\bibinfo {author} {\bibfnamefont {D.}~\bibnamefont
  {Tamascelli}},\ }\bibfield  {title} {\bibinfo {title} {Excitation dynamics in
  chain-mapped environments},\ }\href {https://doi.org/10.3390/e22111320}
  {\bibfield  {journal} {\bibinfo  {journal} {Entropy}\ }\textbf {\bibinfo
  {volume} {22}},\ \bibinfo {pages} {1320} (\bibinfo {year}
  {2020})}\BibitemShut {NoStop}%
\bibitem [{\citenamefont {Dunnett}\ and\ \citenamefont
  {Chin}(2021{\natexlab{a}})}]{Dunnett_Chin_2021_evolving}%
  \BibitemOpen
  \bibfield  {author} {\bibinfo {author} {\bibfnamefont {A.~J.}\ \bibnamefont
  {Dunnett}}\ and\ \bibinfo {author} {\bibfnamefont {A.~W.}\ \bibnamefont
  {Chin}},\ }\bibfield  {title} {\bibinfo {title} {Efficient bond-adaptive
  approach for finite-temperature open quantum dynamics using the one-site
  time-dependent variational principle for matrix product states},\ }\href
  {https://doi.org/10.1103/PhysRevB.104.214302} {\bibfield  {journal} {\bibinfo
   {journal} {Physical Review B}\ }\textbf {\bibinfo {volume} {104}},\ \bibinfo
  {pages} {214302} (\bibinfo {year} {2021}{\natexlab{a}})}\BibitemShut
  {NoStop}%
\bibitem [{\citenamefont {Dunnett}\ and\ \citenamefont
  {Chin}(2021{\natexlab{b}})}]{Dunnett_Chin_2021_vibr}%
  \BibitemOpen
  \bibfield  {author} {\bibinfo {author} {\bibfnamefont {A.~J.}\ \bibnamefont
  {Dunnett}}\ and\ \bibinfo {author} {\bibfnamefont {A.~W.}\ \bibnamefont
  {Chin}},\ }\bibfield  {title} {\bibinfo {title} {Simulating quantum vibronic
  dynamics at finite temperatures with many body wave functions at 0 k},\
  }\href {https://doi.org/10.3389/fchem.2020.600731} {\bibfield  {journal}
  {\bibinfo  {journal} {Frontiers in Chemistry}\ }\textbf {\bibinfo {volume}
  {8}},\ \bibinfo {pages} {1195} (\bibinfo {year}
  {2021}{\natexlab{b}})}\BibitemShut {NoStop}%
\bibitem [{\citenamefont {Dunnett}\ and\ \citenamefont
  {Chin}(2021{\natexlab{c}})}]{Dunnett_Chin_2021}%
  \BibitemOpen
  \bibfield  {author} {\bibinfo {author} {\bibfnamefont {A.~J.}\ \bibnamefont
  {Dunnett}}\ and\ \bibinfo {author} {\bibfnamefont {A.~W.}\ \bibnamefont
  {Chin}},\ }\bibfield  {title} {\bibinfo {title} {Matrix product state
  simulations of non-equilibrium steady states and transient heat flows in the
  two-bath spin-boson model at finite temperatures},\ }\href
  {https://doi.org/10.3390/e23010077} {\bibfield  {journal} {\bibinfo
  {journal} {Entropy}\ }\textbf {\bibinfo {volume} {23}},\ \bibinfo {pages}
  {77} (\bibinfo {year} {2021}{\natexlab{c}})}\BibitemShut {NoStop}%
\bibitem [{\citenamefont {de~Vega}\ and\ \citenamefont
  {Bañuls}(2015)}]{devegaban}%
  \BibitemOpen
  \bibfield  {author} {\bibinfo {author} {\bibfnamefont {I.}~\bibnamefont
  {de~Vega}}\ and\ \bibinfo {author} {\bibfnamefont {M.~C.}\ \bibnamefont
  {Bañuls}},\ }\bibfield  {title} {\bibinfo {title} {Thermofield-based
  chain-mapping approach for open quantum systems},\ }\bibfield  {journal}
  {\bibinfo  {journal} {Physical Review A}\ }\textbf {\bibinfo {volume} {92}},\
  \href {https://doi.org/10.1103/physreva.92.052116}
  {10.1103/physreva.92.052116} (\bibinfo {year} {2015})\BibitemShut {NoStop}%
\bibitem [{\citenamefont {Smirne}\ \emph {et~al.}(2022)\citenamefont {Smirne},
  \citenamefont {Tamascelli}, \citenamefont {Lim}, \citenamefont {Plenio},\
  and\ \citenamefont {Huelga}}]{tamaSmirne22}%
  \BibitemOpen
  \bibfield  {author} {\bibinfo {author} {\bibfnamefont {A.}~\bibnamefont
  {Smirne}}, \bibinfo {author} {\bibfnamefont {D.}~\bibnamefont {Tamascelli}},
  \bibinfo {author} {\bibfnamefont {J.}~\bibnamefont {Lim}}, \bibinfo {author}
  {\bibfnamefont {M.~B.}\ \bibnamefont {Plenio}},\ and\ \bibinfo {author}
  {\bibfnamefont {S.~F.}\ \bibnamefont {Huelga}},\ }\bibfield  {title}
  {\bibinfo {title} {Non-perturbative treatment of open-system multi-time
  expectation values in {G}aussian bosonic environments},\ }\href
  {https://doi.org/10.1142/S1230161222500196} {\bibfield  {journal} {\bibinfo
  {journal} {Open Systems and Information Dynamics}\ }\textbf {\bibinfo
  {volume} {29}},\ \bibinfo {pages} {2250019} (\bibinfo {year}
  {2022})}\BibitemShut {NoStop}%
\bibitem [{\citenamefont {N\"u\ss{}eler}\ \emph {et~al.}(2022)\citenamefont
  {N\"u\ss{}eler}, \citenamefont {Tamascelli}, \citenamefont {Smirne},
  \citenamefont {Lim}, \citenamefont {Huelga},\ and\ \citenamefont
  {Plenio}}]{closure}%
  \BibitemOpen
  \bibfield  {author} {\bibinfo {author} {\bibfnamefont {A.}~\bibnamefont
  {N\"u\ss{}eler}}, \bibinfo {author} {\bibfnamefont {D.}~\bibnamefont
  {Tamascelli}}, \bibinfo {author} {\bibfnamefont {A.}~\bibnamefont {Smirne}},
  \bibinfo {author} {\bibfnamefont {J.}~\bibnamefont {Lim}}, \bibinfo {author}
  {\bibfnamefont {S.~F.}\ \bibnamefont {Huelga}},\ and\ \bibinfo {author}
  {\bibfnamefont {M.~B.}\ \bibnamefont {Plenio}},\ }\bibfield  {title}
  {\bibinfo {title} {Fingerprint and universal markovian closure of structured
  bosonic environments},\ }\href
  {https://doi.org/10.1103/PhysRevLett.129.140604} {\bibfield  {journal}
  {\bibinfo  {journal} {Phys. Rev. Lett.}\ }\textbf {\bibinfo {volume} {129}},\
  \bibinfo {pages} {140604} (\bibinfo {year} {2022})}\BibitemShut {NoStop}%
\bibitem [{\citenamefont {Breuer}\ \emph {et~al.}(2016)\citenamefont {Breuer},
  \citenamefont {Laine}, \citenamefont {Piilo},\ and\ \citenamefont
  {Vacchini}}]{breuer_vacchini}%
  \BibitemOpen
  \bibfield  {author} {\bibinfo {author} {\bibfnamefont {H.~P.}\ \bibnamefont
  {Breuer}}, \bibinfo {author} {\bibfnamefont {E.~M.}\ \bibnamefont {Laine}},
  \bibinfo {author} {\bibfnamefont {J.}~\bibnamefont {Piilo}},\ and\ \bibinfo
  {author} {\bibfnamefont {B.}~\bibnamefont {Vacchini}},\ }\bibfield  {title}
  {\bibinfo {title} {Colloquium: Non-markovian dynamics in open quantum
  systems},\ }\bibfield  {journal} {\bibinfo  {journal} {Reviews of Modern
  Physics}\ }\textbf {\bibinfo {volume} {88}},\ \href
  {https://doi.org/10.1103/revmodphys.88.021002} {10.1103/revmodphys.88.021002}
  (\bibinfo {year} {2016})\BibitemShut {NoStop}%
\bibitem [{\citenamefont {Weiss}(2012)}]{WeissUlrich}%
  \BibitemOpen
  \bibfield  {author} {\bibinfo {author} {\bibfnamefont {U.}~\bibnamefont
  {Weiss}},\ }\href@noop {} {\emph {\bibinfo {title} {Quantum dissipative
  systems}}},\ \bibinfo {edition} {4th}\ ed.\ (\bibinfo  {publisher} {World
  Scientific},\ \bibinfo {address} {Singapore Hackensack},\ \bibinfo {year}
  {2012})\BibitemShut {NoStop}%
\bibitem [{\citenamefont {Feynman}\ and\ \citenamefont
  {Vernon}(1963)}]{FEYNMAN1963118}%
  \BibitemOpen
  \bibfield  {author} {\bibinfo {author} {\bibfnamefont {R.}~\bibnamefont
  {Feynman}}\ and\ \bibinfo {author} {\bibfnamefont {F.}~\bibnamefont
  {Vernon}},\ }\bibfield  {title} {\bibinfo {title} {The theory of a general
  quantum system interacting with a linear dissipative system},\ }\href
  {https://doi.org/https://doi.org/10.1016/0003-4916(63)90068-X} {\bibfield
  {journal} {\bibinfo  {journal} {Annals of Physics}\ }\textbf {\bibinfo
  {volume} {24}},\ \bibinfo {pages} {118} (\bibinfo {year} {1963})}\BibitemShut
  {NoStop}%
\bibitem [{\citenamefont {Chin}\ \emph {et~al.}(2010)\citenamefont {Chin},
  \citenamefont {Rivas}, \citenamefont {Huelga},\ and\ \citenamefont
  {Plenio}}]{chain_mapping_original}%
  \BibitemOpen
  \bibfield  {author} {\bibinfo {author} {\bibfnamefont {A.~W.}\ \bibnamefont
  {Chin}}, \bibinfo {author} {\bibfnamefont {A.}~\bibnamefont {Rivas}},
  \bibinfo {author} {\bibfnamefont {S.~F.}\ \bibnamefont {Huelga}},\ and\
  \bibinfo {author} {\bibfnamefont {M.~B.}\ \bibnamefont {Plenio}},\ }\bibfield
   {title} {\bibinfo {title} {Exact mapping between system-reservoir quantum
  models and semi-infinite discrete chains using orthogonal polynomials},\
  }\href {https://doi.org/10.1063/1.3490188} {\bibfield  {journal} {\bibinfo
  {journal} {Journal of Mathematical Physics}\ }\textbf {\bibinfo {volume}
  {51}},\ \bibinfo {pages} {092109} (\bibinfo {year} {2010})},\ \Eprint
  {https://arxiv.org/abs/https://doi.org/10.1063/1.3490188}
  {https://doi.org/10.1063/1.3490188} \BibitemShut {NoStop}%
\bibitem [{\citenamefont {Woods}\ \emph {et~al.}(2014)\citenamefont {Woods},
  \citenamefont {Groux}, \citenamefont {Chin}, \citenamefont {Huelga},\ and\
  \citenamefont {Plenio}}]{WoodsChin}%
  \BibitemOpen
  \bibfield  {author} {\bibinfo {author} {\bibfnamefont {M.~P.}\ \bibnamefont
  {Woods}}, \bibinfo {author} {\bibfnamefont {R.}~\bibnamefont {Groux}},
  \bibinfo {author} {\bibfnamefont {A.~W.}\ \bibnamefont {Chin}}, \bibinfo
  {author} {\bibfnamefont {S.~F.}\ \bibnamefont {Huelga}},\ and\ \bibinfo
  {author} {\bibfnamefont {M.~B.}\ \bibnamefont {Plenio}},\ }\bibfield  {title}
  {\bibinfo {title} {Mappings of open quantum systems onto chain
  representations and markovian embeddings},\ }\href
  {https://doi.org/10.1063/1.4866769} {\bibfield  {journal} {\bibinfo
  {journal} {Journal of Mathematical Physics}\ }\textbf {\bibinfo {volume}
  {55}},\ \bibinfo {pages} {032101} (\bibinfo {year} {2014})}\BibitemShut
  {NoStop}%
\bibitem [{\citenamefont {Gautschi}(1994)}]{orthpol}%
  \BibitemOpen
  \bibfield  {author} {\bibinfo {author} {\bibfnamefont {W.}~\bibnamefont
  {Gautschi}},\ }\bibfield  {title} {\bibinfo {title} {Algorithm 726:
  Orthpol–a package of routines for generating orthogonal polynomials and
  gauss-type quadrature rules},\ }\href {https://doi.org/10.1145/174603.174605}
  {\bibfield  {journal} {\bibinfo  {journal} {ACM Trans. Math. Softw.}\
  }\textbf {\bibinfo {volume} {20}},\ \bibinfo {pages} {21–62} (\bibinfo
  {year} {1994})}\BibitemShut {NoStop}%
\bibitem [{Note1()}]{Note1}%
  \BibitemOpen
  \bibinfo {note} {The numerical methods employed use the software packages
  available at: \protect \url
  {https://github.com/tfmlaX/Chaincoeffs}}\BibitemShut {NoStop}%
\bibitem [{\citenamefont {Haroche}\ and\ \citenamefont
  {Raimond}(2006)}]{Haroche:993568}%
  \BibitemOpen
  \bibfield  {author} {\bibinfo {author} {\bibfnamefont {S.}~\bibnamefont
  {Haroche}}\ and\ \bibinfo {author} {\bibfnamefont {J.~M.}\ \bibnamefont
  {Raimond}},\ }\href
  {https://doi.org/10.1093/acprof:oso/9780198509141.001.0001} {\emph {\bibinfo
  {title} {{Exploring the Quantum: Atoms, Cavities, and Photons}}}}\ (\bibinfo
  {publisher} {Oxford Univ. Press},\ \bibinfo {address} {Oxford},\ \bibinfo
  {year} {2006})\BibitemShut {NoStop}%
\bibitem [{\citenamefont {Lacroix}\ \emph {et~al.}(2021)\citenamefont
  {Lacroix}, \citenamefont {Dunnett}, \citenamefont {Gribben}, \citenamefont
  {Lovett},\ and\ \citenamefont {Chin}}]{lacroix2021unveiling}%
  \BibitemOpen
  \bibfield  {author} {\bibinfo {author} {\bibfnamefont {T.}~\bibnamefont
  {Lacroix}}, \bibinfo {author} {\bibfnamefont {A.}~\bibnamefont {Dunnett}},
  \bibinfo {author} {\bibfnamefont {D.}~\bibnamefont {Gribben}}, \bibinfo
  {author} {\bibfnamefont {B.~W.}\ \bibnamefont {Lovett}},\ and\ \bibinfo
  {author} {\bibfnamefont {A.}~\bibnamefont {Chin}},\ }\bibfield  {title}
  {\bibinfo {title} {Unveiling non-markovian spacetime signaling in open
  quantum systems with long-range tensor network dynamics},\ }\href@noop {}
  {\bibfield  {journal} {\bibinfo  {journal} {Physical Review A}\ }\textbf
  {\bibinfo {volume} {104}},\ \bibinfo {pages} {052204} (\bibinfo {year}
  {2021})}\BibitemShut {NoStop}%
\bibitem [{\citenamefont {Chin}\ \emph {et~al.}(2011)\citenamefont {Chin},
  \citenamefont {Prior}, \citenamefont {Huelga},\ and\ \citenamefont
  {Plenio}}]{Chin_QPT_sub-ohmic_SBModel_ansatz}%
  \BibitemOpen
  \bibfield  {author} {\bibinfo {author} {\bibfnamefont {A.~W.}\ \bibnamefont
  {Chin}}, \bibinfo {author} {\bibfnamefont {J.}~\bibnamefont {Prior}},
  \bibinfo {author} {\bibfnamefont {S.~F.}\ \bibnamefont {Huelga}},\ and\
  \bibinfo {author} {\bibfnamefont {M.~B.}\ \bibnamefont {Plenio}},\ }\bibfield
   {title} {\bibinfo {title} {Generalized polaron ansatz for the ground state
  of the sub-ohmic spin-boson model: An analytic theory of the localization
  transition},\ }\href {https://doi.org/10.1103/PhysRevLett.107.160601}
  {\bibfield  {journal} {\bibinfo  {journal} {Physical Review Letters}\
  }\textbf {\bibinfo {volume} {107}},\ \bibinfo {pages} {160601} (\bibinfo
  {year} {2011})}\BibitemShut {NoStop}%
\bibitem [{\citenamefont {Irish}\ and\ \citenamefont
  {Gea-Banacloche}(2014)}]{Irish_Gea-Banacloche_2014}%
  \BibitemOpen
  \bibfield  {author} {\bibinfo {author} {\bibfnamefont {E.~K.}\ \bibnamefont
  {Irish}}\ and\ \bibinfo {author} {\bibfnamefont {J.}~\bibnamefont
  {Gea-Banacloche}},\ }\bibfield  {title} {\bibinfo {title} {Oscillator
  tunneling dynamics in the rabi model},\ }\href
  {https://doi.org/10.1103/PhysRevB.89.085421} {\bibfield  {journal} {\bibinfo
  {journal} {Physical Review B}\ }\textbf {\bibinfo {volume} {89}},\ \bibinfo
  {pages} {085421} (\bibinfo {year} {2014})},\ \bibinfo {note} {arXiv:
  1312.5121}\BibitemShut {NoStop}%
\bibitem [{\citenamefont {Martinazzo}\ \emph {et~al.}(2011)\citenamefont
  {Martinazzo}, \citenamefont {Vacchini}, \citenamefont {Hughes},\ and\
  \citenamefont {Burghardt}}]{martinazzo2011communication}%
  \BibitemOpen
  \bibfield  {author} {\bibinfo {author} {\bibfnamefont {R.}~\bibnamefont
  {Martinazzo}}, \bibinfo {author} {\bibfnamefont {B.}~\bibnamefont
  {Vacchini}}, \bibinfo {author} {\bibfnamefont {K.~H.}\ \bibnamefont
  {Hughes}},\ and\ \bibinfo {author} {\bibfnamefont {I.}~\bibnamefont
  {Burghardt}},\ }\bibfield  {title} {\bibinfo {title} {Communication:
  Universal markovian reduction of brownian particle dynamics},\ }\href@noop {}
  {\bibfield  {journal} {\bibinfo  {journal} {The Journal of Chemical Physics}\
  }\textbf {\bibinfo {volume} {134}},\ \bibinfo {pages} {011101} (\bibinfo
  {year} {2011})}\BibitemShut {NoStop}%
\bibitem [{\citenamefont {Iles-Smith}\ \emph {et~al.}(2016)\citenamefont
  {Iles-Smith}, \citenamefont {Dijkstra}, \citenamefont {Lambert},\ and\
  \citenamefont {Nazir}}]{iles2016energy}%
  \BibitemOpen
  \bibfield  {author} {\bibinfo {author} {\bibfnamefont {J.}~\bibnamefont
  {Iles-Smith}}, \bibinfo {author} {\bibfnamefont {A.~G.}\ \bibnamefont
  {Dijkstra}}, \bibinfo {author} {\bibfnamefont {N.}~\bibnamefont {Lambert}},\
  and\ \bibinfo {author} {\bibfnamefont {A.}~\bibnamefont {Nazir}},\ }\bibfield
   {title} {\bibinfo {title} {Energy transfer in structured and unstructured
  environments: Master equations beyond the born-markov approximations},\
  }\href@noop {} {\bibfield  {journal} {\bibinfo  {journal} {The Journal of
  chemical physics}\ }\textbf {\bibinfo {volume} {144}},\ \bibinfo {pages}
  {044110} (\bibinfo {year} {2016})}\BibitemShut {NoStop}%
\bibitem [{\citenamefont {Nazir}\ and\ \citenamefont
  {Schaller}(2018{\natexlab{a}})}]{nazir2018reaction}%
  \BibitemOpen
  \bibfield  {author} {\bibinfo {author} {\bibfnamefont {A.}~\bibnamefont
  {Nazir}}\ and\ \bibinfo {author} {\bibfnamefont {G.}~\bibnamefont
  {Schaller}},\ }\bibfield  {title} {\bibinfo {title} {The reaction coordinate
  mapping in quantum thermodynamics},\ }\href@noop {} {\bibfield  {journal}
  {\bibinfo  {journal} {Thermodynamics in the Quantum Regime: Fundamental
  Aspects and New Directions}\ ,\ \bibinfo {pages} {551}} (\bibinfo {year}
  {2018}{\natexlab{a}})}\BibitemShut {NoStop}%
\bibitem [{\citenamefont {Blunden-Codd}\ \emph {et~al.}(2017)\citenamefont
  {Blunden-Codd}, \citenamefont {Bera}, \citenamefont {Bruognolo},
  \citenamefont {Linden}, \citenamefont {Chin}, \citenamefont {von Delft},
  \citenamefont {Nazir},\ and\ \citenamefont {Florens}}]{blunden2017anatomy}%
  \BibitemOpen
  \bibfield  {author} {\bibinfo {author} {\bibfnamefont {Z.}~\bibnamefont
  {Blunden-Codd}}, \bibinfo {author} {\bibfnamefont {S.}~\bibnamefont {Bera}},
  \bibinfo {author} {\bibfnamefont {B.}~\bibnamefont {Bruognolo}}, \bibinfo
  {author} {\bibfnamefont {N.-O.}\ \bibnamefont {Linden}}, \bibinfo {author}
  {\bibfnamefont {A.~W.}\ \bibnamefont {Chin}}, \bibinfo {author}
  {\bibfnamefont {J.}~\bibnamefont {von Delft}}, \bibinfo {author}
  {\bibfnamefont {A.}~\bibnamefont {Nazir}},\ and\ \bibinfo {author}
  {\bibfnamefont {S.}~\bibnamefont {Florens}},\ }\bibfield  {title} {\bibinfo
  {title} {Anatomy of quantum critical wave functions in dissipative impurity
  problems},\ }\href@noop {} {\bibfield  {journal} {\bibinfo  {journal}
  {Physical Review B}\ }\textbf {\bibinfo {volume} {95}},\ \bibinfo {pages}
  {085104} (\bibinfo {year} {2017})}\BibitemShut {NoStop}%
\bibitem [{\citenamefont {Nacke}\ \emph {et~al.}(2023)\citenamefont {Nacke},
  \citenamefont {Otterpohl}, \citenamefont {Thorwart},\ and\ \citenamefont
  {Nalbach}}]{nacke23}%
  \BibitemOpen
  \bibfield  {author} {\bibinfo {author} {\bibfnamefont {P.}~\bibnamefont
  {Nacke}}, \bibinfo {author} {\bibfnamefont {F.}~\bibnamefont {Otterpohl}},
  \bibinfo {author} {\bibfnamefont {M.}~\bibnamefont {Thorwart}},\ and\
  \bibinfo {author} {\bibfnamefont {P.}~\bibnamefont {Nalbach}},\ }\bibfield
  {title} {\bibinfo {title} {Dephasing and pseudocoherent quantum dynamics in
  super-ohmic environments},\ }\href
  {https://doi.org/10.1103/PhysRevA.107.062218} {\bibfield  {journal} {\bibinfo
   {journal} {Phys. Rev. A}\ }\textbf {\bibinfo {volume} {107}},\ \bibinfo
  {pages} {062218} (\bibinfo {year} {2023})}\BibitemShut {NoStop}%
\bibitem [{\citenamefont {Metha}\ \emph {et~al.}(2016)\citenamefont {Metha},
  \citenamefont {Bruzewicz}, \citenamefont {McConnell}, \citenamefont {Ram},
  \citenamefont {Sage},\ and\ \citenamefont {Chiaverini}}]{metha16}%
  \BibitemOpen
  \bibfield  {author} {\bibinfo {author} {\bibfnamefont {K.~K.}\ \bibnamefont
  {Metha}}, \bibinfo {author} {\bibfnamefont {C.~D.}\ \bibnamefont
  {Bruzewicz}}, \bibinfo {author} {\bibfnamefont {R.}~\bibnamefont
  {McConnell}}, \bibinfo {author} {\bibfnamefont {R.~J.}\ \bibnamefont {Ram}},
  \bibinfo {author} {\bibfnamefont {R.~J.}\ \bibnamefont {Sage}},\ and\
  \bibinfo {author} {\bibfnamefont {J.}~\bibnamefont {Chiaverini}},\ }\bibfield
   {title} {\bibinfo {title} {Integrated optical addressing of an ion qubit},\
  }\href {https://doi.org/10.1038/nnano.2016.139} {\bibfield  {journal}
  {\bibinfo  {journal} {Nat. Nano.}\ }\textbf {\bibinfo {volume} {11}},\
  \bibinfo {pages} {1066} (\bibinfo {year} {2016})}\BibitemShut {NoStop}%
\bibitem [{\citenamefont {Gambetta}\ \emph {et~al.}(2017)\citenamefont
  {Gambetta}, \citenamefont {Chow},\ and\ \citenamefont
  {Steffen}}]{gambetta17}%
  \BibitemOpen
  \bibfield  {author} {\bibinfo {author} {\bibfnamefont {J.~M.}\ \bibnamefont
  {Gambetta}}, \bibinfo {author} {\bibfnamefont {J.~M.}\ \bibnamefont {Chow}},\
  and\ \bibinfo {author} {\bibfnamefont {M.}~\bibnamefont {Steffen}},\
  }\bibfield  {title} {\bibinfo {title} {Building logical qubits in a
  superconducting quantum computing system},\ }\href
  {https://doi.org/10.1038/s41534-016-0004-0} {\bibfield  {journal} {\bibinfo
  {journal} {npj,Quantum Inf.}\ }\textbf {\bibinfo {volume} {3}},\ \bibinfo
  {pages} {2} (\bibinfo {year} {2017})}\BibitemShut {NoStop}%
\bibitem [{\citenamefont {Esposito}\ \emph {et~al.}(2015)\citenamefont
  {Esposito}, \citenamefont {Ochoa},\ and\ \citenamefont
  {Galperin}}]{esposito15}%
  \BibitemOpen
  \bibfield  {author} {\bibinfo {author} {\bibfnamefont {M.}~\bibnamefont
  {Esposito}}, \bibinfo {author} {\bibfnamefont {M.~A.}\ \bibnamefont
  {Ochoa}},\ and\ \bibinfo {author} {\bibfnamefont {M.}~\bibnamefont
  {Galperin}},\ }\bibfield  {title} {\bibinfo {title} {Quantum thermodynamics:
  A nonequilibrium green's function approach},\ }\href
  {https://doi.org/10.1103/PhysRevLett.114.080602} {\bibfield  {journal}
  {\bibinfo  {journal} {Phys. Rev. Lett.}\ }\textbf {\bibinfo {volume} {114}},\
  \bibinfo {pages} {080602} (\bibinfo {year} {2015})}\BibitemShut {NoStop}%
\bibitem [{\citenamefont {Purkayastha}\ \emph {et~al.}(2022)\citenamefont
  {Purkayastha}, \citenamefont {Guarnieri}, \citenamefont {Campbell},
  \citenamefont {Prior},\ and\ \citenamefont {Goold}}]{prior22}%
  \BibitemOpen
  \bibfield  {author} {\bibinfo {author} {\bibfnamefont {A.}~\bibnamefont
  {Purkayastha}}, \bibinfo {author} {\bibfnamefont {G.}~\bibnamefont
  {Guarnieri}}, \bibinfo {author} {\bibfnamefont {S.}~\bibnamefont {Campbell}},
  \bibinfo {author} {\bibfnamefont {J.}~\bibnamefont {Prior}},\ and\ \bibinfo
  {author} {\bibfnamefont {J.}~\bibnamefont {Goold}},\ }\bibfield  {title}
  {\bibinfo {title} {Periodically refreshed quantum thermal machines},\ }\href
  {https://doi.org/10.22331/q-2022-09-08-801} {\bibfield  {journal} {\bibinfo
  {journal} {{Quantum}}\ }\textbf {\bibinfo {volume} {6}},\ \bibinfo {pages}
  {801} (\bibinfo {year} {2022})}\BibitemShut {NoStop}%
\bibitem [{\citenamefont {Smirne}\ \emph {et~al.}(2016)\citenamefont {Smirne},
  \citenamefont {Ko\l{}ody\ifmmode~\acute{n}\else \'{n}\fi{}ski}, \citenamefont
  {Huelga},\ and\ \citenamefont {Demkowicz-Dobrza\ifmmode~\acute{n}\else
  \'{n}\fi{}ski}}]{smirne16}%
  \BibitemOpen
  \bibfield  {author} {\bibinfo {author} {\bibfnamefont {A.}~\bibnamefont
  {Smirne}}, \bibinfo {author} {\bibfnamefont {J.}~\bibnamefont
  {Ko\l{}ody\ifmmode~\acute{n}\else \'{n}\fi{}ski}}, \bibinfo {author}
  {\bibfnamefont {S.~F.}\ \bibnamefont {Huelga}},\ and\ \bibinfo {author}
  {\bibfnamefont {R.}~\bibnamefont {Demkowicz-Dobrza\ifmmode~\acute{n}\else
  \'{n}\fi{}ski}},\ }\bibfield  {title} {\bibinfo {title} {Ultimate precision
  limits for noisy frequency estimation},\ }\href
  {https://doi.org/10.1103/PhysRevLett.116.120801} {\bibfield  {journal}
  {\bibinfo  {journal} {Phys. Rev. Lett.}\ }\textbf {\bibinfo {volume} {116}},\
  \bibinfo {pages} {120801} (\bibinfo {year} {2016})}\BibitemShut {NoStop}%
\bibitem [{\citenamefont {Ribeiro}\ and\ \citenamefont
  {Vieira}(2015)}]{ribeiro15}%
  \BibitemOpen
  \bibfield  {author} {\bibinfo {author} {\bibfnamefont {P.}~\bibnamefont
  {Ribeiro}}\ and\ \bibinfo {author} {\bibfnamefont {V.~R.}\ \bibnamefont
  {Vieira}},\ }\bibfield  {title} {\bibinfo {title} {Non-markovian effects in
  electronic and spin transport},\ }\href
  {https://doi.org/10.1103/PhysRevB.92.100302} {\bibfield  {journal} {\bibinfo
  {journal} {Phys. Rev. B}\ }\textbf {\bibinfo {volume} {92}},\ \bibinfo
  {pages} {100302} (\bibinfo {year} {2015})}\BibitemShut {NoStop}%
\bibitem [{\citenamefont {Mitchison}\ and\ \citenamefont
  {Plenio}(2018)}]{mitchison18}%
  \BibitemOpen
  \bibfield  {author} {\bibinfo {author} {\bibfnamefont {M.~T.}\ \bibnamefont
  {Mitchison}}\ and\ \bibinfo {author} {\bibfnamefont {M.~B.}\ \bibnamefont
  {Plenio}},\ }\bibfield  {title} {\bibinfo {title} {Non-additive dissipation
  in open quantum networks out of equilibrium},\ }\href
  {https://doi.org/10.1088/1367-2630/aa9f70} {\bibfield  {journal} {\bibinfo
  {journal} {New Journal of Physics}\ }\textbf {\bibinfo {volume} {20}},\
  \bibinfo {pages} {033005} (\bibinfo {year} {2018})}\BibitemShut {NoStop}%
\bibitem [{\citenamefont {Huelga}\ and\ \citenamefont
  {Plenio}(2013)}]{huelga13}%
  \BibitemOpen
  \bibfield  {author} {\bibinfo {author} {\bibfnamefont {S.~F.}\ \bibnamefont
  {Huelga}}\ and\ \bibinfo {author} {\bibfnamefont {M.~B.}\ \bibnamefont
  {Plenio}},\ }\bibfield  {title} {\bibinfo {title} {Vibrations, quanta and
  biology},\ }\href {https://doi.org/10.1080/00405000.2013.829687} {\bibfield
  {journal} {\bibinfo  {journal} {Contemp. Phys.}\ }\textbf {\bibinfo {volume}
  {54}},\ \bibinfo {pages} {181} (\bibinfo {year} {2013})},\ \Eprint
  {https://arxiv.org/abs/https://doi.org/10.1080/00405000.2013.829687}
  {https://doi.org/10.1080/00405000.2013.829687} \BibitemShut {NoStop}%
\bibitem [{\citenamefont {Caycedo-Soler}\ \emph {et~al.}(2022)\citenamefont
  {Caycedo-Soler}, \citenamefont {Mattioni}, \citenamefont {Lim}, \citenamefont
  {Renger}, \citenamefont {Hulega},\ and\ \citenamefont {Plenio}}]{felipe22}%
  \BibitemOpen
  \bibfield  {author} {\bibinfo {author} {\bibfnamefont {F.}~\bibnamefont
  {Caycedo-Soler}}, \bibinfo {author} {\bibfnamefont {A.}~\bibnamefont
  {Mattioni}}, \bibinfo {author} {\bibfnamefont {J.}~\bibnamefont {Lim}},
  \bibinfo {author} {\bibfnamefont {T.}~\bibnamefont {Renger}}, \bibinfo
  {author} {\bibfnamefont {S.~F.}\ \bibnamefont {Hulega}},\ and\ \bibinfo
  {author} {\bibfnamefont {M.~B.}\ \bibnamefont {Plenio}},\ }\bibfield  {title}
  {\bibinfo {title} {Exact simulation of pigment-protein complexes unveils
  vibronic renormalization of electronic parameters in ultrafast
  spectroscopy},\ }\href {https://doi.org/10.1038/s41467-022-30565-4}
  {\bibfield  {journal} {\bibinfo  {journal} {Nat. Commun.}\ }\textbf {\bibinfo
  {volume} {13}},\ \bibinfo {pages} {2912} (\bibinfo {year}
  {2022})}\BibitemShut {NoStop}%
\bibitem [{\citenamefont {Nazir}\ and\ \citenamefont
  {Schaller}(2018{\natexlab{b}})}]{nazir18}%
  \BibitemOpen
  \bibfield  {author} {\bibinfo {author} {\bibfnamefont {A.}~\bibnamefont
  {Nazir}}\ and\ \bibinfo {author} {\bibfnamefont {G.}~\bibnamefont
  {Schaller}},\ }\bibfield  {title} {\bibinfo {title} {The reaction coordinate
  mapping in quantum thermodynamic},\ }in\ \href@noop {} {\emph {\bibinfo
  {booktitle} {Thermodynamics in the Quantum Regime. Fundamental Theories in
  Physics, vol. 195}}},\ \bibinfo {editor} {edited by\ \bibinfo {editor}
  {\bibfnamefont {F.}~\bibnamefont {Binder}}, \bibinfo {editor} {\bibfnamefont
  {L.}~\bibnamefont {Correa}}, \bibinfo {editor} {\bibfnamefont
  {C.}~\bibnamefont {Gogolin}}, \bibinfo {editor} {\bibfnamefont
  {J.}~\bibnamefont {Anders}},\ and\ \bibinfo {editor} {\bibfnamefont
  {G.}~\bibnamefont {Adesso}}}\ (\bibinfo  {publisher} {Springer, Cham},\
  \bibinfo {address} {Switzerland AG},\ \bibinfo {year} {2018})\BibitemShut
  {NoStop}%
\bibitem [{\citenamefont {Lorenzoni}\ \emph {et~al.}(2023)\citenamefont
  {Lorenzoni}, \citenamefont {Cho}, \citenamefont {Lim}, \citenamefont
  {Tamascelli}, \citenamefont {Huelga},\ and\ \citenamefont
  {Plenio}}]{lorenzoni23}%
  \BibitemOpen
  \bibfield  {author} {\bibinfo {author} {\bibfnamefont {N.}~\bibnamefont
  {Lorenzoni}}, \bibinfo {author} {\bibfnamefont {N.}~\bibnamefont {Cho}},
  \bibinfo {author} {\bibfnamefont {J.}~\bibnamefont {Lim}}, \bibinfo {author}
  {\bibfnamefont {D.}~\bibnamefont {Tamascelli}}, \bibinfo {author}
  {\bibfnamefont {S.~F.}\ \bibnamefont {Huelga}},\ and\ \bibinfo {author}
  {\bibfnamefont {M.~B.}\ \bibnamefont {Plenio}},\ }\href@noop {} {\bibinfo
  {title} {Systematic coarse-graining of environments for the non-perturbative
  simulation of open quantum systems}} (\bibinfo {year} {2023}),\ \Eprint
  {https://arxiv.org/abs/2303.08982} {arXiv:2303.08982 [quant-ph]} \BibitemShut
  {NoStop}%
\bibitem [{\citenamefont {Tamascelli}\ \emph {et~al.}(2018)\citenamefont
  {Tamascelli}, \citenamefont {Smirne}, \citenamefont {Huelga},\ and\
  \citenamefont {Plenio}}]{tama18}%
  \BibitemOpen
  \bibfield  {author} {\bibinfo {author} {\bibfnamefont {D.}~\bibnamefont
  {Tamascelli}}, \bibinfo {author} {\bibfnamefont {A.}~\bibnamefont {Smirne}},
  \bibinfo {author} {\bibfnamefont {S.~F.}\ \bibnamefont {Huelga}},\ and\
  \bibinfo {author} {\bibfnamefont {M.~B.}\ \bibnamefont {Plenio}},\ }\bibfield
   {title} {\bibinfo {title} {Nonperturbative treatment of non-markovian
  dynamics of open quantum systems},\ }\href
  {https://doi.org/10.1103/PhysRevLett.120.030402} {\bibfield  {journal}
  {\bibinfo  {journal} {Phys. Rev. Lett.}\ }\textbf {\bibinfo {volume} {120}},\
  \bibinfo {pages} {030402} (\bibinfo {year} {2018})}\BibitemShut {NoStop}%
\bibitem [{\citenamefont {Mascherpa}\ \emph {et~al.}(2020)\citenamefont
  {Mascherpa}, \citenamefont {Smirne}, \citenamefont {Somoza}, \citenamefont
  {Fern\'andez-Acebal}, \citenamefont {Donadi}, \citenamefont {Tamascelli},
  \citenamefont {Huelga},\ and\ \citenamefont {Plenio}}]{mascherpa20}%
  \BibitemOpen
  \bibfield  {author} {\bibinfo {author} {\bibfnamefont {F.}~\bibnamefont
  {Mascherpa}}, \bibinfo {author} {\bibfnamefont {A.}~\bibnamefont {Smirne}},
  \bibinfo {author} {\bibfnamefont {A.~D.}\ \bibnamefont {Somoza}}, \bibinfo
  {author} {\bibfnamefont {P.}~\bibnamefont {Fern\'andez-Acebal}}, \bibinfo
  {author} {\bibfnamefont {S.}~\bibnamefont {Donadi}}, \bibinfo {author}
  {\bibfnamefont {D.}~\bibnamefont {Tamascelli}}, \bibinfo {author}
  {\bibfnamefont {S.~F.}\ \bibnamefont {Huelga}},\ and\ \bibinfo {author}
  {\bibfnamefont {M.~B.}\ \bibnamefont {Plenio}},\ }\bibfield  {title}
  {\bibinfo {title} {Optimized auxiliary oscillators for the simulation of
  general open quantum systems},\ }\href
  {https://doi.org/10.1103/PhysRevA.101.052108} {\bibfield  {journal} {\bibinfo
   {journal} {Phys. Rev. A}\ }\textbf {\bibinfo {volume} {101}},\ \bibinfo
  {pages} {052108} (\bibinfo {year} {2020})}\BibitemShut {NoStop}%
\bibitem [{\citenamefont {Metelmann}\ and\ \citenamefont
  {Clerk}(2015)}]{metelmann}%
  \BibitemOpen
  \bibfield  {author} {\bibinfo {author} {\bibfnamefont {A.}~\bibnamefont
  {Metelmann}}\ and\ \bibinfo {author} {\bibfnamefont {A.~A.}\ \bibnamefont
  {Clerk}},\ }\bibfield  {title} {\bibinfo {title} {Nonreciprocal photon
  transmission and amplification via reservoir engineering},\ }\href
  {https://doi.org/10.1103/PhysRevX.5.021025} {\bibfield  {journal} {\bibinfo
  {journal} {Phys. Rev. X}\ }\textbf {\bibinfo {volume} {5}},\ \bibinfo {pages}
  {021025} (\bibinfo {year} {2015})}\BibitemShut {NoStop}%
\bibitem [{\citenamefont {Mirrahimi}\ \emph {et~al.}(2014)\citenamefont
  {Mirrahimi}, \citenamefont {Leghtas}, \citenamefont {Albert}, \citenamefont
  {Touzard}, \citenamefont {Schoelkopf}, \citenamefont {Jiang},\ and\
  \citenamefont {Devoret}}]{Mirrahimi_2014}%
  \BibitemOpen
  \bibfield  {author} {\bibinfo {author} {\bibfnamefont {M.}~\bibnamefont
  {Mirrahimi}}, \bibinfo {author} {\bibfnamefont {Z.}~\bibnamefont {Leghtas}},
  \bibinfo {author} {\bibfnamefont {V.~V.}\ \bibnamefont {Albert}}, \bibinfo
  {author} {\bibfnamefont {S.}~\bibnamefont {Touzard}}, \bibinfo {author}
  {\bibfnamefont {R.~J.}\ \bibnamefont {Schoelkopf}}, \bibinfo {author}
  {\bibfnamefont {L.}~\bibnamefont {Jiang}},\ and\ \bibinfo {author}
  {\bibfnamefont {M.~H.}\ \bibnamefont {Devoret}},\ }\bibfield  {title}
  {\bibinfo {title} {Dynamically protected cat-qubits: a new paradigm for
  universal quantum computation},\ }\href
  {https://doi.org/10.1088/1367-2630/16/4/045014} {\bibfield  {journal}
  {\bibinfo  {journal} {New Journal of Physics}\ }\textbf {\bibinfo {volume}
  {16}},\ \bibinfo {pages} {045014} (\bibinfo {year} {2014})}\BibitemShut
  {NoStop}%
\bibitem [{\citenamefont {Cirac}\ \emph {et~al.}(2021)\citenamefont {Cirac},
  \citenamefont {Perez-Garcia}, \citenamefont {Schuch},\ and\ \citenamefont
  {Verstraete}}]{Cirac_Perez-Garcia_Schuch_Verstraete_2021}%
  \BibitemOpen
  \bibfield  {author} {\bibinfo {author} {\bibfnamefont {I.}~\bibnamefont
  {Cirac}}, \bibinfo {author} {\bibfnamefont {D.}~\bibnamefont {Perez-Garcia}},
  \bibinfo {author} {\bibfnamefont {N.}~\bibnamefont {Schuch}},\ and\ \bibinfo
  {author} {\bibfnamefont {F.}~\bibnamefont {Verstraete}},\ }\bibfield  {title}
  {\bibinfo {title} {Matrix product states and projected entangled pair states:
  Concepts, symmetries, and theorems},\ }\href
  {http://arxiv.org/abs/2011.12127} {\bibfield  {journal} {\bibinfo  {journal}
  {arXiv:2011.12127 [cond-mat, physics:hep-th, physics:quant-ph]}\ } (\bibinfo
  {year} {2021})},\ \bibinfo {note} {arXiv: 2011.12127}\BibitemShut {NoStop}%
\bibitem [{\citenamefont {Dunnett}\ and\ \citenamefont
  {Chin}(2020)}]{dunnett2020dynamically}%
  \BibitemOpen
  \bibfield  {author} {\bibinfo {author} {\bibfnamefont {A.~J.}\ \bibnamefont
  {Dunnett}}\ and\ \bibinfo {author} {\bibfnamefont {A.~W.}\ \bibnamefont
  {Chin}},\ }\href@noop {} {\bibinfo {title} {Dynamically evolving
  bond-dimensions within the one-site time-dependent-variational-principle
  method for matrix product states: Towards efficient simulation of
  non-equilibrium open quantum dynamics}} (\bibinfo {year} {2020}),\ \Eprint
  {https://arxiv.org/abs/2007.13528} {arXiv:2007.13528 [quant-ph]} \BibitemShut
  {NoStop}%
\bibitem [{Note2()}]{Note2}%
  \BibitemOpen
  \bibinfo {note} {The numerical methods employed use the software packages
  available at: \protect \url
  {https://github.com/angusdunnett/MPSDynamics}.}\BibitemShut {Stop}%
\end{thebibliography}
\end{document}